\documentclass[twocolumn]{aastex702}

\usepackage{amsmath}
\usepackage{amssymb}
\usepackage{graphicx}
\definecolor{trackchange}{rgb}{0.75,0.0,0.0}
\newcommand{\rev}[1]{#1}

\newcommand{\beq}{\begin{equation}}
\newcommand{\eeq}{\end{equation}}
\newcommand{\Msun}{M_{\odot}}
\newcommand{\Porb}{P_{\rm orb}}
\newcommand{\Pref}{P_{\rm ref}}
\newcommand{\Teff}{T_{\rm eff}}
\newcommand{\logg}{\log g}
\newcommand{\MG}{M_G}
\newcommand{\fGW}{f_{\rm GW}}
\newcommand{\Mc}{\mathcal{M}_c}

\shorttitle{Short-Period Blue Compact-Binary Candidates}
\shortauthors{Lin et al.}

\begin{document}

\title{A Systematic Gaia--ZTF Search for Short-Period Blue Compact-Binary Candidates}

\author[0009-0008-9942-620X]{Jiamao Lin}
\email{linjm66@mail2.sysu.edu.cn}
\affiliation{School of Physics and Astronomy, Sun Yat-sen University, Zhuhai 519082, China}
\affiliation{CSST Science Center for the Guangdong-Hongkong-Macau Greater Bay Area, Sun Yat-sen University, Zhuhai 519082, China}

\author[0000-0002-1428-4003]{Liangliang Ren}
\email{rll@ahstu.edu.cn}
\correspondingauthor{Liangliang Ren}
\affiliation{School of Electrical and Electronic Engineering, Anhui Science and Technology University, Bengbu, Anhui 233030, China}

\author{Yilong Li}
\email{liyl69@mail2.sysu.edu.cn}
\affiliation{School of Physics and Astronomy, Sun Yat-sen University, Zhuhai 519082, China}

\author[0000-0002-0378-2023]{Bo Ma}
\email{mabo8@mail.sysu.edu.cn}
\affiliation{School of Physics and Astronomy, Sun Yat-sen University, Zhuhai 519082, China}
\affiliation{CSST Science Center for the Guangdong-Hongkong-Macau Greater Bay Area, Sun Yat-sen University, Zhuhai 519082, China}

\author[0000-0003-0707-3213]{Di-Chang Chen}
\email{chendch28@mail.sysu.edu.cn}
\affiliation{School of Physics and Astronomy, Sun Yat-sen University, Zhuhai 519082, China}
\affiliation{CSST Science Center for the Guangdong-Hongkong-Macau Greater Bay Area, Sun Yat-sen University, Zhuhai 519082, China}

\author{Zi-Heng Yu}
\email{}
\affiliation{School of Science, Shenzhen Campus of Sun Yat-sen University, Shenzhen 518107, China}

\author{Sen Yang}
\email{}
\affiliation{School of Science, Shenzhen Campus of Sun Yat-sen University, Shenzhen 518107, China}

\author[0000-0002-7112-759X]{Shun-Jia Huang}
\email{huangshj69@sysu.edu.cn}
\affiliation{School of Science, Shenzhen Campus of Sun Yat-sen University, Shenzhen 518107, China}

\author[0000-0002-7869-0174]{Yi-Ming Hu}
\email{huyiming@mail.sysu.edu.cn}
\affiliation{School of Physics and Astronomy, Sun Yat-sen University, Zhuhai 519082, China}
\affiliation{MOE Key Laboratory of TianQin Mission, TianQin Research Center for Gravitational Physics, Frontiers Science Center for TianQin, Gravitational Wave Research Center of CNSA, Sun Yat-sen University, Zhuhai 519082, China}

\author[0000-0002-3084-5157]{Chengyuan Li}
\email{lichengy5@mail.sysu.edu.cn}
\affiliation{School of Physics and Astronomy, Sun Yat-sen University, Zhuhai 519082, China}
\affiliation{CSST Science Center for the Guangdong-Hongkong-Macau Greater Bay Area, Sun Yat-sen University, Zhuhai 519082, China}

\begin{abstract}
Short-period binaries containing white dwarfs or hot subdwarfs are among the strongest Galactic sources for millihertz gravitational-wave (GW) observatories, but the confirmed sample remains small. Combining Gaia DR3 astrometry and photometry with ZTF DR23 light curves---through a Gaia selection calibrated against known systems, ZTF period searches, and machine-learning morphology ranking---we assemble a catalog of 147 short-period (10.34--106.46~min) blue compact-binary candidates, 111 of them without a previous compact-binary classification. DESI DR1, GALEX, and AllWISE data reveal an intrinsically heterogeneous sample: on the Gaia color--magnitude diagram it comprises 52 white-dwarf-locus, 69 hot-subdwarf-region, and 26 intermediate sources, and among the 26 members with DESI spectra only about one-third lie on the white-dwarf cooling sequence, the rest being more luminous blue stars whose low-resolution spectra resemble white dwarfs. We identify a prioritized set of new short-period white-dwarf-locus candidates for time-resolved follow-up, including ten with periods below 40~min and none with an existing radial-velocity series. Under fiducial binary assumptions, 17 newly identified white-dwarf-locus candidates would exceed the adopted LISA threshold, led by a 37~pc white dwarf; \added{no candidate has a measured chirp mass, and with the signal-to-noise ratio scaling as $\Mc^{5/3}$ the count is 9 for $\Mc=0.15\,\Msun$, 17 for the adopted $0.3\,\Msun$, and 21 for $0.6\,\Msun$, in each case assuming the modulation is orbital.} We show, however, that for the compact majority of the white-dwarf-locus sample the observed modulation amplitudes exceed any possible ellipsoidal signal by three to five orders of magnitude, so a rotating magnetic or chemically inhomogeneous single white dwarf is a competing interpretation that ZTF photometry alone cannot exclude; the catalog contains at least one confirmed case. We release the full 147-source catalog with periods, Gaia and spectroscopic classifications, harmonic and ellipsoidal diagnostics, and supplementary tables of fiducial GW estimates and ultraviolet--infrared photometry.
\end{abstract}


\section{Introduction}
\label{sec:intro}

Short-period binaries containing white dwarfs or hot subdwarfs are a major Galactic source population for the planned millihertz gravitational-wave (GW) observatories TianQin \citep{Luo2016} and LISA \citep{LISA2017, AmaroSeoane2023}. The relevant classes include detached double white dwarfs (DWDs), semidetached AM~CVn systems, hot subdwarf (sdB/sdO) binaries, cataclysmic variables, and related ultracompact systems \citep{Kupfer2018, Kupfer2024}. Many---but not all---of these channels involve common-envelope evolution and subsequent orbital decay, and together they probe mass transfer and angular-momentum loss \citep{Iben1984, Webbink1984, Nelemans2001, Ivanova2013}. Their brightest members include the LISA/TianQin ``verification binaries,'' while the population as a whole contributes to the mHz Galactic GW foreground \citep{Kupfer2018, Huang2020}. At favorable inclinations, eclipses, ellipsoidal modulation, reflection effects, or accretion-related variability can reveal these systems photometrically \citep{Hermes2012, Burdge2019a}. Double white dwarfs are a primary target of this search, but a photometric selection based on blue, short-period variability necessarily returns a broader mix of compact-binary candidates.

Population synthesis predicts $\sim 10^8$ DWDs in the Milky Way, with $\sim 10^4$ at $\Porb < 1$~hr and $\fGW \gtrsim 0.5$~mHz \citep{Nelemans2001, Lamberts2019, AmaroSeoane2023}, yet the confirmed short-period compact binaries with spectroscopic orbital solutions remain a small sample: the most recent Gaia-informed verification-binary compilation lists only $\sim$40 sources detectable over the LISA mission \citep{Kupfer2024}. This census gap is primarily observational: the systems are intrinsically faint ($G \gtrsim 17$~mag), require time-series photometry to reveal their periodicity, and need substantial spectroscopy to confirm. The gap is widest outside the DA double-white-dwarf channel, because radial-velocity confirmation relies on the strong Balmer lines that DA white dwarfs provide, leaving line-poor DB/DC white dwarfs, AM~CVn systems, and hot-subdwarf binaries comparatively unexplored. The known sample remains too small to calibrate GW population models for TianQin and LISA mission planning \citep{Nissanke2012, AmaroSeoane2023}.

Wide-field time-domain surveys have transformed the photometric side of this census. The productive strategy is a systematic color-and-variability selection applied to a full survey: \citet{Burdge2020a} searched color-selected ZTF light curves and identified 15 ultracompact LISA-detectable binaries with orbital periods of 7--56~min, seven of them eclipsing double white dwarfs, and \citet{Ren2023} combined Gaia EDR3 color--magnitude selection with ZTF DR8 variability and manual light-curve inspection to assemble the largest catalog of close white-dwarf binary (CWDB) candidates so far. The individual systems these searches deliver set the current state of the art: ZTF~J1539+5027, a $\Porb=6.91$~min eclipsing DWD whose GW-driven orbital decay is measurable within a few years \citep{Burdge2019a}; the 8.8-min eclipsing detached double white dwarf ZTF~J2243+5242 \citep{Burdge2020b}; the 20.5-min detached ellipsoidal binary TMTS/ZTF~J0526+5934, whose visible component is interpreted as either a compact helium-burning hot subdwarf \citep{Lin2024} or an extremely low-mass white dwarf \citep{Rebassa2024a}; and the Roche-lobe-filling sdB+WD binaries ZTF~J2130+4420 (39~min) and ZTF~J2055+4651 (56~min) \citep{Kupfer2020}, which confirm that variability searches recover the full range of compact-binary types, not double white dwarfs alone.

Spectroscopic surveys provide a complementary, inclination-independent route, and have followed the same progression from dedicated programs to survey scale. The ELM Survey compiled over a hundred extremely low-mass WD binaries with orbital solutions \citep{Kilic2021}, and the MUCHFUSS project systematically measured the short-period hot-subdwarf binary population \citep{Kupfer2015}. Massively multiplexed spectrographs now apply radial-velocity selection to millions of targets: \citet{Jiang2025} identified 33 DA DWD candidates, 28 of them new, from DESI EDR spectra, and \citet{Pallathadka2025} identified 60 DA DWD candidates, 43 of them new, from SDSS-V DR19 multi-epoch velocities, with tentative periods for 9 systems. Radial-velocity selection yields stronger direct binary constraints than single-epoch spectra or photometric modulation alone.

Despite this progress, three limitations persist. Selection thresholds are still commonly chosen by eye, so the selection function of a published catalog cannot be reconstructed; light-curve vetting remains largely manual and therefore does not scale to LSST \citep{Ivezic2019} or CSST \citep{Gong2019}; and only a minority of photometric candidates have any spectrum, so the astrophysical mix that a blue, short-period photometric cut actually admits is assumed rather than measured. The third limitation is now removable: the DESI DR1 Milky Way Survey \citep{DESIDR1_2025} supplies mid-resolution spectra for millions of stars over a footprint that overlaps ZTF.

We therefore carried out a Gaia--ZTF search designed to address all three, and this defines what the present work adds to earlier Gaia--ZTF catalogs. Every Gaia threshold is fixed by a quantitative retention-rate ratio measured against a literature reference set, so the selection is reproducible from the published numbers alone. Light-curve vetting is performed by a trained network rather than by inspection, so the procedure transfers directly to LSST- and CSST-scale inputs. Finally, DESI DR1 spectra, ultraviolet and infrared photometry, and Gaia luminosities are used to characterize the resulting sample, which turns out to be intrinsically heterogeneous---a result that a purely photometric catalog cannot establish about itself.

Operationally the search runs in four steps: five calibrated Gaia DR3 cuts; period searches of the surviving ZTF DR23 light curves; morphology ranking of the phase-folded light curves with a convolutional neural network (CNN) based on the MobileNetV2 architecture \citep{Sandler2018}; and a Box Least Squares (BLS) refinement, applied to every ranked target, that resolves the harmonic and alias ambiguities of the detection-stage periodogram before a source enters the catalog. Gaia color--magnitude positions, DESI DR1 spectra, ultraviolet and infrared photometry, and fiducial gravitational-wave estimates then characterize the sample. The search yields 147 short-period blue compact-binary candidates, 111 of them without a previous compact-binary classification.

This paper is organized as follows. Section~\ref{sec:data} describes the reference catalogs and survey data. Section~\ref{sec:method} details the Gaia filtering, period searches, morphology ranking, period refinement, and DESI cross-match. Section~\ref{sec:results} presents the catalog, CMD distribution, and spectroscopic characterization. Section~\ref{sec:discussion} compares earlier searches and gives fiducial gravitational-wave and multiwavelength interpretations. Section~\ref{sec:summary} summarizes the results.

\section{Data}
\label{sec:data}

This section describes only the data the search is built from: the literature reference set used to calibrate the Gaia thresholds, the two surveys that supply the astrometry and the light curves, the spectroscopic survey used for characterization, and the ancillary archives. How these data are filtered, searched, and cross-matched is deferred to Section~\ref{sec:method}.

\subsection{Known SPWDB Reference Catalog}
\label{sec:known}

We assembled a literature-compiled reference set of short-period white-dwarf binaries (SPWDBs) and candidates, drawn from \citet{Burdge2020b}, \citet{Hermes2012}, \citet{Burdge2019b}, \citet{Chandra2021}, \citet{Kosakowski2023}, \citet{Ren2023}, \citet{Burdge2020a}, and \citet{Munday2023}. \added{After cross-matching with Gaia DR3 (3~arcsec radius) and removing duplicate rows, 209 unique Gaia sources remain.}

This set has a single purpose: to mark where in Gaia parameter space real short-period white-dwarf binaries actually lie. Instead of placing selection boundaries by inspection, we can then ask of any candidate threshold how many of these 209 systems it preserves for a given loss of field stars, and adopt the value that maximizes the contrast (Section~\ref{sec:gaia_filter}). Two properties limit how far the set can be pushed. First, it is heterogeneous---spectroscopically confirmed binaries and purely photometric candidates are mixed together---so it is a calibration sample rather than a truth set, and the fraction of it that survives the pipeline measures how the pipeline treats already-known systems, not the completeness of any physical binary class. Second, it is disjoint from the labeled light-curve images used to train the morphology-ranking network (Section~\ref{sec:ml_samples}), so no object contributes both to the Gaia thresholds and to the network weights.

\subsection{Gaia DR3}
\label{sec:gaia_data}

Gaia Data Release 3 \citep{GaiaDR32022} provides positions, parallaxes, proper motions, and three-band photometry ($G$, $G_{\rm BP}$, $G_{\rm RP}$) for $\sim$1.8~billion sources observed from 2014 July to 2017 May. Three of its properties are what make a catalog-level preselection of this population possible: all-sky coverage, parallaxes precise enough to place a faint blue star on or off the white-dwarf cooling sequence, and homogeneous colors on a single photometric system. Gaia also reports per-source photometric scatter and astrometric goodness-of-fit statistics, which carry information on variability and on blending independently of any light curve. We accessed Gaia DR3 through the Gaia Archive ADQL interface.

\subsection{ZTF Light Curves}
\label{sec:ztf}

The Zwicky Transient Facility \citep{Bellm2019, Graham2019, Masci2019} is a time-domain survey operating on the Palomar 48-inch Schmidt telescope, covering the northern sky ($\delta > -28^\circ$) in $g$, $r$, and $i$ bands to $5\sigma$ limiting magnitudes of $\sim$20.8, 20.6, and 20.2~mag, respectively. Its 47~deg$^2$ camera and 30~s exposures deliver a survey whose sampling is sparse and irregular but whose baseline is long: we use ZTF Data Release 23, whose public-survey epochs span up to $\approx$6.6~yr ($\sim$2400~d). This combination governs what a minute-timescale search can and cannot do. A 30~s exposure already spans 5\% of a 10-min period, so the shortest signals are partly smeared by the integration itself; the nightly and seasonal sampling imprints strong aliases at the sidereal day and at the survey cadence; and the multi-year baseline allows a coherent period, once identified, to be refined to a precision far beyond the nominal exposure timescale.

\subsection{DESI DR1 Milky Way Survey}
\label{sec:desi_data}

The Dark Energy Spectroscopic Instrument \citep{DESI2016} is a multi-object spectrograph on the 4-m Mayall telescope whose 5000 robotic fiber positioners cover a 3\fdg2 field. Its DR1 release \citep{DESIDR1_2025} provides mid-resolution ($R\sim2700$) optical spectroscopy over $\lambda\approx3600$--9800~\AA. The Milky Way Survey component includes extensive coverage of faint blue stellar sources \citep{Koposov2026}, which is why a blue, short-period photometric catalog has appreciable spectroscopic overlap at all. We use the calibrated B-, R-, and Z-arm spectra; DESI pipeline radial velocities and atmospheric parameters are not used in this work. The spectra are single-epoch coadds, so they carry line and continuum information but no orbital velocity curve.

\subsection{Ancillary Archives}
\label{sec:ancillary}

Ultraviolet and infrared photometry for the multiwavelength characterization of Appendix~\ref{app:uvir} is drawn from GALEX (FUV, NUV), SDSS, Pan-STARRS, 2MASS, and AllWISE; among the WISE bands only $W1$ and $W2$ reach these targets, while $W3$ and $W4$ are too shallow. Published spectral types, object types, and prior variability classifications are taken from SIMBAD and VizieR. For the release version used here, Gaia DR3 astrometry and photometry were queried from the Gaia Archive through ADQL, ZTF DR23 light curves were obtained from the public ZTF data-release service, and DESI DR1 spectra and Milky Way Survey value-added products were taken from the public DESI DR1 archive; the SIMBAD, VizieR, MAST/GALEX, Pan-STARRS, 2MASS, and AllWISE cross-matches were re-verified on 2026 July 2--3.

\section{Methodology}
\label{sec:method}

The task is to reduce $1.8\times10^{9}$ Gaia sources to a list short enough to examine one object at a time, and the order of the steps follows from their cost. Each stage below is more expensive per source than the one before it, and is therefore applied to fewer sources. Catalog-level Gaia cuts require no new data and can be evaluated across the entire archive, so they run first and carry the bulk of the reduction. Period searching needs a full time series per source and is applied only to the $\sim$10$^{5}$ survivors. Judging a folded light curve is more expensive still and does not scale to that many objects, so it is delegated to a trained network, used to rank rather than to classify. The per-source period refinement, refolding, and manual examination on which the final catalog rests are reserved for the few hundred highest-ranked targets.

The ordering within the Gaia stage is likewise deliberate. The color--magnitude cut comes first because it is the only criterion that defines the population of interest; the quality, kinematic, and astrometric criteria that follow are refinements whose thresholds are meaningful only for sources already in the right region of the diagram. The variability criterion is applied last because it is by far the lossiest of the five and is the one cut that deliberately trades completeness for a tractable number of light curves, so isolating it at the end keeps that cost visible and quantifiable. The criteria are independent per-source conditions, so a different order would return the same final sample; what it would change is the stage-by-stage yields of Table~\ref{tab:gaia_criteria}, which is why those yields are reported against a fixed denominator.

\subsection{Overview of the Pipeline}
\label{sec:overview}

\begin{figure*}[htbp]
\centering
\includegraphics[width=0.92\textwidth]{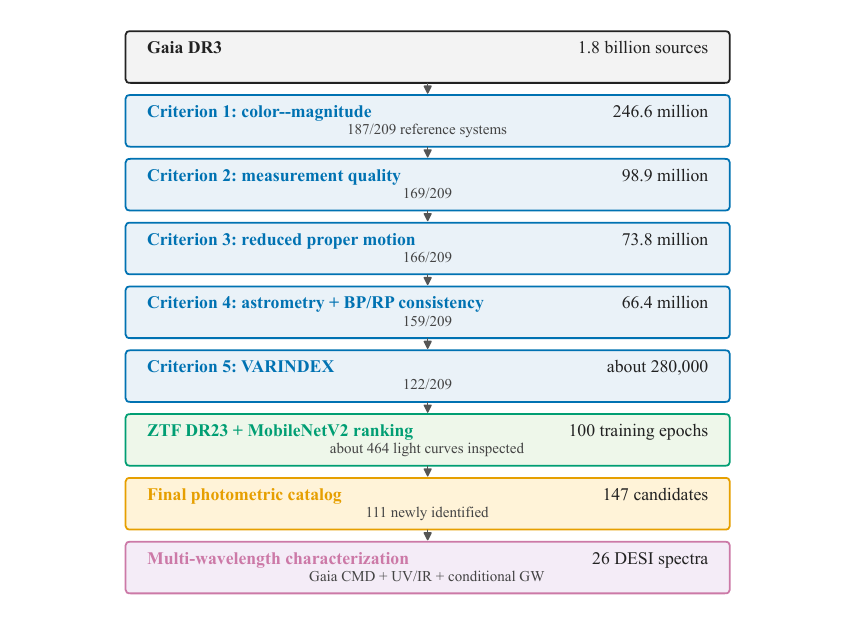}
\caption{Candidate-selection pipeline. The five Gaia DR3 criteria (retentions of the 209-source reference set annotated) reduce the field to $\sim$280,000 ZTF targets; CNN ranking yields 464 promoted targets, and per-source period refinement and refolding produce the final catalog of 147 candidates (111 new; 26 with DESI DR1 spectra).}
\label{fig:pipeline}
\end{figure*}

Figure~\ref{fig:pipeline} summarizes four stages of the search: (1) five sequential Gaia DR3 cuts; (2) ZTF DR23 period searches; (3) CNN morphology ranking followed by high-precision period refinement and refolding of every ranked light curve; and (4) multiwavelength characterization of the final catalog. The CNN uses the MobileNetV2 architecture and is employed only to prioritize EA/EW-like folded morphologies; it does not determine physical binary class or final catalog membership.

\subsection{Gaia DR3 Multi-Dimensional Filtering}
\label{sec:gaia_filter}
\label{sec:selection_function}

The five criteria below are calibrated to maximize retention of the 209 known SPWDBs relative to the field-star background. They select blue, subluminous Gaia sources with short-period variability---a region of parameter space occupied by detached DWDs, AM~CVn systems, hot-subdwarf binaries, cataclysmic variables, and rotationally variable white dwarfs alike \citep{Kupfer2018,Burdge2020a,Ren2023}, so all of these classes are expected in the catalog. All sources satisfying the initial color--magnitude cut of Section~\ref{sec:cmd} were retrieved from the Gaia Archive together with the photometric and astrometric quantities used by the remaining criteria. We apply the five cuts in sequence. For each cut we sweep a grid of thresholds and maximize
\beq
R_{\rm ratio}(\theta) = \frac{R_{\rm known}(\theta)}{R_{\rm Gaia}(\theta)},
\eeq
where $R_{\rm known}(\theta)$ and $R_{\rm Gaia}(\theta)$ are the fractional retentions of known SPWDBs and Gaia background sources at threshold $\theta$. Table~\ref{tab:gaia_criteria} lists the adopted thresholds and stage yields; equations and diagnostic figures for the quality, kinematic, astrometric, and variability cuts appear in Appendix~\ref{app:gaia_criteria}.

\subsubsection{Color--Magnitude Diagram Cut}
\label{sec:cmd}

SPWDBs occupy a distinctive locus in the Hertzsprung--Russell diagram, lying below the main sequence at relatively blue colors. Motivated by standard blue compact-object color--magnitude selections and calibrated below, we select sources satisfying:
\beq
\MG > 3.7\,(G_{\rm BP}-G_{\rm RP}) + 2.4
\label{eq:cmd1}
\eeq
where $\MG = G + 5\log_{10}(\varpi/100)$ is the absolute $G$-band magnitude and $\varpi$ is the positive Gaia parallax in mas. We additionally require:
\beq
G_{\rm BP} - G_{\rm RP} < 1.5
\label{eq:cmd2}
\eeq
following \citet{Ren2023}, who showed that blue selection effectively rejects red main-sequence interlopers while retaining the bulk of the SPWDB locus. Application to the full Gaia DR3 catalog retains $\sim$$2.5\times10^{8}$ sources while preserving 187/209 (89.5\%) of the known SPWDBs.

\subsubsection{Quality, Kinematics, Astrometry, and Variability Cuts}
\label{sec:quality}
\label{sec:rpm}
\label{sec:uwe}
\label{sec:varindex}

The next four cuts remove saturated or low-S/N photometry, apply a reduced-proper-motion criterion, enforce unit-weight-error (UWE) and BP/RP flux-excess limits, and keep sources with variability above the photon-noise floor ($\texttt{VARINDEX}>0$). Starting from the 209-source reference set, the five criteria retain 187, 169, 166, 159, and 122 sources in sequence, i.e.\ 89.5, 80.9, 79.4, 76.1, and 58.4\% of the reference set. The field population is reduced to approximately 66.4 million sources after Criterion~4 and to approximately 280,000 sources after the final variability cut.

\begin{deluxetable}{clcc}
\tablewidth{0pt}
\tabletypesize{\footnotesize}
\tablecaption{Gaia DR3 selection stages\label{tab:gaia_criteria}}
\tablehead{
  \colhead{Stage} & \colhead{Adopted cut (summary)} & \colhead{Cumulative} & \colhead{Stage} \\
  \colhead{} & \colhead{} & \colhead{(of 209)} & \colhead{(of survivors)}
}
\startdata
CMD & $\MG > 3.7(G_{\rm BP}-G_{\rm RP})+2.4$, $G_{\rm BP}-G_{\rm RP}<1.5$ & 187 (89.5\%) & 89.5\% \\
Quality & Parallax/photometry S/N and bright-limit cuts & 169 (80.9\%) & 90.4\% \\
Reduced $H_G$ & $H_G > 5.3(G_{\rm BP}-G_{\rm RP})+5.9$ & 166 (79.4\%) & 98.2\% \\
UWE + $E$ & Lindegren UWE limit; Pelisoli BP/RP excess band & 159 (76.1\%) & 95.8\% \\
VARINDEX & $\texttt{VARINDEX}>0$ (above photon-noise floor) & 122 (58.4\%) & 76.7\%
\enddata
\tablecomments{Retention of the 209-source reference set. The cumulative column is referred to the full reference set throughout and is the quantity to compare between stages; the stage column, referred to the survivors of the previous cut, is retained only because it identifies which individual criterion is costly---the criteria are independent per-source conditions, so the final sample does not depend on the order in which they are applied. Full equations and diagnostic figures appear in Appendix~\ref{app:gaia_criteria}. After Criterion~5 the Gaia pool is $\sim$280{,}000 sources.}
\end{deluxetable}

\subsection{Period Search and Adopted Orbital-period Estimates}
\label{sec:mhaov}
\label{sec:period_def}

We cross-match the $\sim$280,000 Gaia-selected candidates (Section~\ref{sec:varindex}) with ZTF DR23 using a 3~arcsec matching radius and keep sources with $\geq 50$ observations in the $g$ or $r$ band, which is the minimum sampling for which the permutation significance below is meaningful. Each surviving light curve is screened with the Multi-Harmonic Analysis of Variance periodogram (MHAOV; \citealt{SchwarzenbergCzerny1996}), as implemented in \texttt{P4J} \citep{P4J2018},\footnote{\url{https://github.com/phuijse/P4J}} which we adopt because its multi-harmonic model is sensitive to the non-sinusoidal, eclipse-like profiles that a purely sinusoidal periodogram suppresses. The detection-stage search covers periods of 3--100~min, with significance assessed by 1000 flux permutations. Significant detections are supplied to the CNN as phase-folded images for morphology ranking.

The MHAOV screening period can lock onto a harmonic or alias of the true repeating period---the detection period of ES~Cet, for example, is twice its published orbital period---so a fold at the raw detection value can smear or distort the light-curve shape. For promoted candidates we therefore refine the signal locally with a morphology-appropriate high-precision search: GPU-accelerated Box Least Squares \citep{Kovacs2002} for eclipse-dominated curves and Lomb--Scargle or phase-dispersion minimization for smooth modulation. The full-precision repeating photometric period used for this refinement and for phase folding is denoted $\Pref$. All 464 ranked targets were refolded at their refined periods, with their periodograms, band-by-band folds, and phase coherence over the ZTF baseline re-examined; the 147 that retain a coherent short-period fold constitute the catalog.

The catalog reports one adopted orbital-period estimate, denoted $\Porb$: the orbital period inferred from the photometric harmonic of $\Pref$ adopted source by source. An EW-like morphology label does not by itself imply a physical contact binary, nor does it automatically set $\Porb=2\Pref$. The 3--100~min search window applies to the detected repeating period $\Pref$, which spans 10.25--99.05~min across the catalog; because a double-wave (EW-like) morphology repeats twice per orbit, the adopted $\Porb$ of the longest-period EW-like systems extends modestly beyond the search boundary, to a maximum of 106.46~min, giving an adopted $\Porb$ range of 10.34--106.46~min. Except where an independent orbit exists, $\Porb$ is a photometric estimate requiring radial-velocity confirmation. Each period carries a conservative uncertainty $\Delta P_{\rm cons}$, defined as the larger of two terms: the local refinement grid spacing ($\leq P^{2}/(20T)$ over the full ZTF baseline $T$) and the spread between the independently refined $g$- and $r$-band periods. For well-sampled sources $\Delta P_{\rm cons}$ is typically below 0.001~min (Tables~\ref{tab:high_priority}, \ref{tab:desi}, and \ref{tab:lconly}); it should be read as a repeatability bound on the photometric period, not as a formal orbital-period error. Three multiband-poor cases (J081638, J145259, and J214140) show band-to-band disagreements of 1.7--6.5~min and correspondingly large $\Delta P_{\rm cons}$, so their adopted periods are unreliable and they do not enter the high-priority or GW rankings.

\subsection{Machine-learning Morphology Ranking}
\label{sec:ml_samples}

\begin{figure*}[htbp]
\centering
\includegraphics[width=0.95\textwidth]{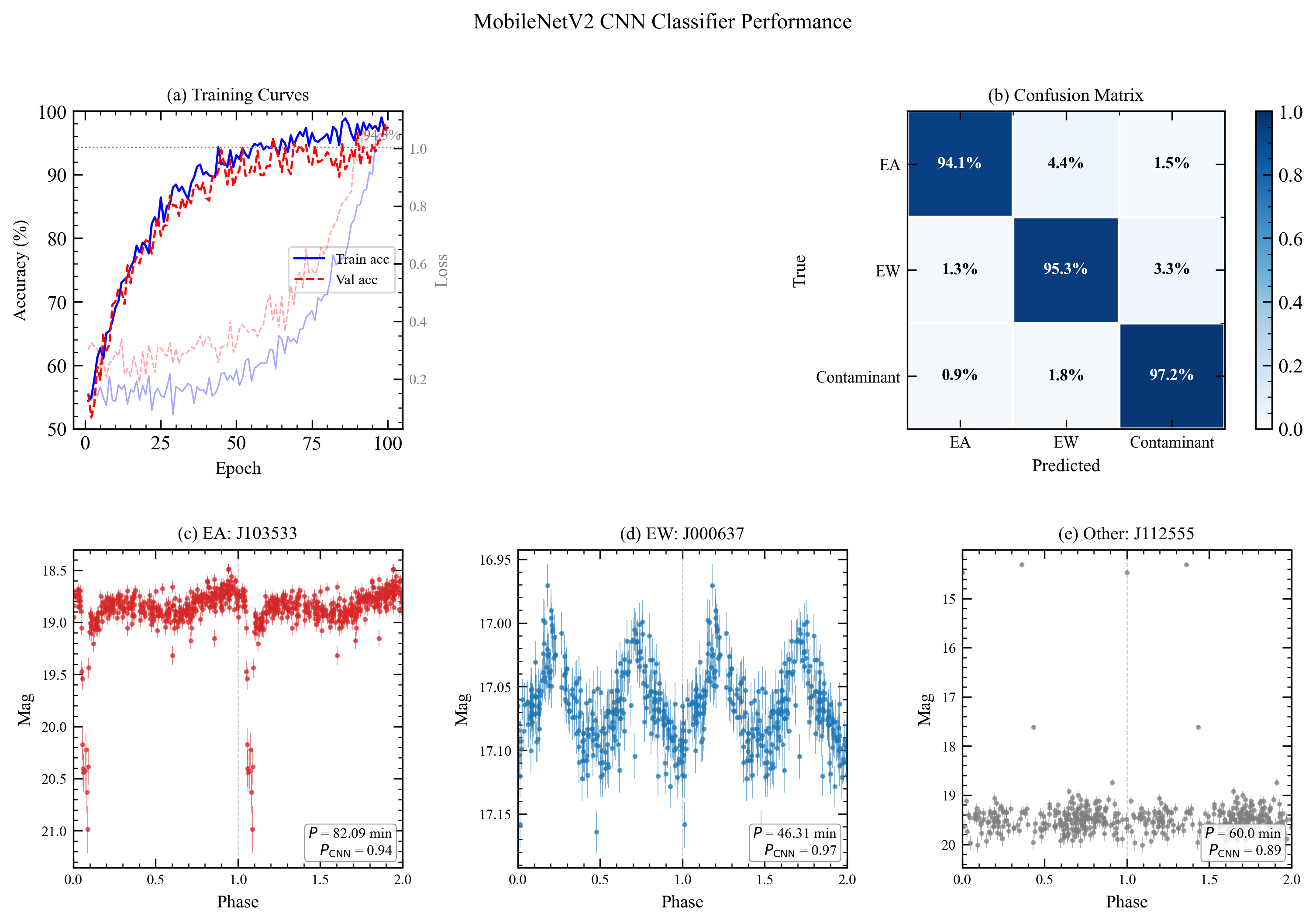}
\caption{CNN morphology-ranking performance using the MobileNetV2 architecture. \textit{(a)} Training and validation accuracy and loss over 100 epochs. \textit{(b)} Row-normalized confusion matrix for the EA, EW, and Other classes. \textit{(c)--(e)} Representative phase-folded light curves from the three classes.}
\label{fig:cnn}
\end{figure*}

We adopt MobileNetV2 \citep{Sandler2018} as the CNN architecture\footnote{Reference implementation: \url{https://github.com/tensorflow/models/tree/master/research/slim/nets/mobilenet}.} and use it to rank folded morphologies, not to classify binaries. MobileNetV2 replaces each standard convolution with a depthwise separable pair---one spatial filter per input channel, followed by a $1\times1$ combination across channels---and stacks these inside inverted residual blocks with linear bottlenecks, so that the wide feature maps live only inside a block while the shortcut connections carry a narrow representation. The result is a network of $\sim$3.4~million parameters, roughly an order of magnitude smaller than the standard deep image classifiers of comparable benchmark accuracy. That is the reason for the choice here. Distinguishing an eclipse-like from a smooth double-wave fold is a low-complexity shape-recognition problem on small, low-information images, for which a compact network trains to convergence on a few thousand examples without the overfitting a larger model would show, while remaining cheap enough to apply to the full set of detection-stage folds. It was trained (94.3\% held-out accuracy; Figure~\ref{fig:cnn})\added{, using the reference implementation cited above without architectural modification,} on $\sim$3{,}000 EA/EW-like and $\sim$5{,}000 other phase-folded images, with all augmented views of a source kept in one split; the three labels (EA-like, EW-like, other) describe morphology only. The training images are folded ZTF light curves of eclipsing and non-eclipsing variables drawn from general variable-star catalogs and labeled by folded appearance, deliberately spanning a wider period range than the search window so that the network learns shape rather than period; they are disjoint from the 209-source reference set used to calibrate the Gaia cuts (Section~\ref{sec:known}).

The 464 promoted targets were refolded at their refined periods and entered the catalog only if they showed a coherent short-period signal, a plausible morphology, and consistency with the Gaia selection.

\subsection{DESI DR1 Spectroscopic Cross-Matching}
\label{sec:desi_method}

Twenty-six members of the final catalog have DESI DR1 counterparts. We cross-match the catalog with the DESI DR1 stellar database using a 3~arcsec radius after propagating Gaia DR3 positions from epoch 2016.0 to the DESI observing epoch. The largest propagated displacement is much smaller than the adopted radius. The analytic DESI stellar surface-density estimate gives an expected total of approximately 0.012 random associations over the 147 positions, so chance matches are negligible in the spectroscopic subset.

For each final source, we retrieve the calibrated B-, R-, and Z-arm spectra. Flux units are $10^{-17}$~erg~s$^{-1}$~cm$^{-2}$~\AA$^{-1}$. For display only, isolated one- to four-pixel cosmic-ray or sky-subtraction spikes are replaced by a local rolling median, with higher rejection thresholds near common Balmer, He, Ca, and Na lines. All line descriptions use the unmodified spectra. Because the available observations do not provide an orbital radial-velocity series, we make no DESI orbital-solution or component-mass claim.

\section{Results}
\label{sec:results}

\subsection{Final Candidate Sample}
\label{sec:sample}

The Gaia--ZTF search yielded 464 CNN-ranked targets, of which 147 passed period refinement and constitute the final catalog, with adopted periods spanning 10.34--106.46~min (Section~\ref{sec:period_def}). Their composition by CMD region, folded morphology, novelty, and DESI coverage is given in Table~\ref{tab:composition}.

The funnel from $\sim$280{,}000 Gaia-selected targets to 147 is steep and not uniformly vetted, and we state its limits explicitly. Most MHAOV detections above the adopted permutation threshold are aliases of the sidereal day or of the survey cadence rather than astrophysical signals; CNN ranking promotes 464 targets, of which 147 survive period refinement. Each of the 147 was refolded and examined, but the 317 rejected promotions were not individually documented, so that step is reproducible only at the level of the criteria in Section~\ref{sec:period_def}. End to end the pipeline recovers 28 of the 209 reference systems (13\%), \added{with the Gaia variability cut, ZTF coverage, and the detection-and-vetting stage each contributing a measurable share of the loss (Section~\ref{sec:completeness})}.

\added{The detection stage searched $\sim$$2.8\times10^{5}$ light curves, each with $\approx$$1.1\times10^{6}$ independent frequencies in the 3--100~min window (median baseline 2400~d, $\Delta f=465.6$~d$^{-1}$), about $3\times10^{11}$ trials in total. The 1000-permutation test resolves false-alarm probabilities only down to $10^{-3}$ per light curve, so several hundred noise triggers are expected among the detections; refinement indeed rejected 317 of the 464 CNN promotions. Catalog membership, however, requires the independently refined $g$- and $r$-band periods to agree, and an unrelated pair of frequencies does so with probability $2\delta f/\Delta f=9\times10^{-7}$ at the median released tolerance. Summing these probabilities over the catalog gives 0.23 expected spurious two-band agreements among the 147, or 0.7 if all 464 refined targets are counted as noise trials, with just under half of either sum in the five sources with the weakest multiband constraints. Consistent with a small false-alarm contribution, 65 of the 70 members whose light curves can be split at the median epoch have half-baseline profile correlations above 0.5 in both bands (medians 0.86 and 0.82, against a noise scatter of $1/\sqrt{29}=0.19$), and 63 of the 147 (43\%) were catalogued as variables or compact binaries independently of this search. The tolerance is measured from the data rather than fixed in advance, and the two bands share a sampling window and hence alias structure, so these are estimates rather than bounds. We expect of order one of the 147 entries to be a period-search artifact, most likely among the multiband-poor minority.}

\begin{figure*}[htbp]
\centering
\includegraphics[width=0.88\textwidth]{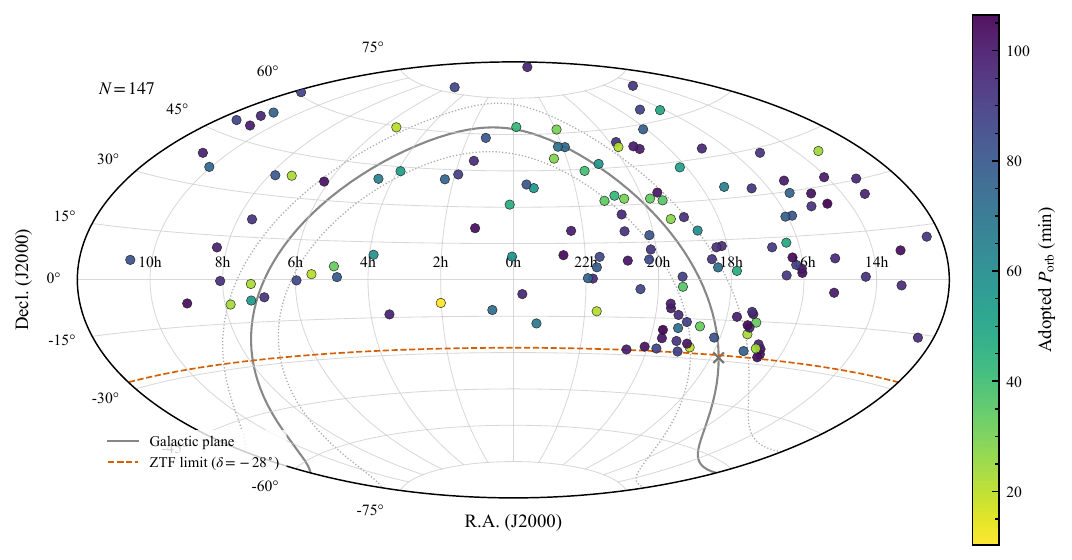}
\caption{Sky distribution of the 147 catalog candidates in an Aitoff projection, with R.A. increasing to the left. All sources are shown as circles and color-coded by the adopted photometric orbital-period estimate $\Porb$. The orange dashed curve marks the nominal ZTF southern boundary at $\delta=-28^\circ$. The solid grey curve is the Galactic plane ($b=0^\circ$) and the dotted grey curves $b=\pm10^\circ$; the cross marks the Galactic center. The latitude distribution relative to this reference is discussed in Section~\ref{sec:spatial_period}.}
\label{fig:sky}
\end{figure*}

Novelty is assigned by cross-matching against the literature reference set, \citet{Ren2023}, SIMBAD, and general variable-star catalogs (Section~\ref{sec:comparison}): most new candidates have no prior variable-star entry, and 21 of the 26 DESI members are characterized spectroscopically here for the first time. Novelty refers to compact-binary classification only: 21 of the 31 newly identified WD-locus members carry a published spectral type, and one has a published rotation period (Sections~\ref{sec:recovered} and \ref{sec:harmonic_test}). Figure~\ref{fig:sky} shows the sky distribution of the full catalog, released in machine-readable form.

\begin{deluxetable}{llc}
\tablewidth{0pt}
\tabletypesize{\footnotesize}
\tablecaption{Composition of the 147-source catalog\label{tab:composition}}
\tablehead{
  \colhead{Category} & \colhead{Subset} & \colhead{$N$}
}
\startdata
Gaia CMD region & white-dwarf locus & 52 \\
 & hot-subdwarf region & 69 \\
 & intermediate & 26 \\
\hline
Folded morphology & EW-like & 130 \\
 & EA-like & 17 \\
\hline
Novelty & newly identified & 111 \\
 & \quad WD locus / hot-sd / interm. & 31 / 59 / 21 \\
 & \quad no prior variable-star entry & 84 \\
 & previously classified & 36 \\
\hline
DESI DR1 spectra & matched members & 26 \\
 & \quad WD-CS / lum.-blue / peculiar & 9 / 14 / 3 \\
\hline
Follow-up subsets & $\MG\geq11$ (26 in WD locus) & 27 \\
 & WD locus, $\Porb<40$~min & 15 \\
 & \quad newly identified & 10 \\
 & \quad + $\MG\geq11$, $\varpi/\sigma_\varpi\geq5$ & 8 \\
 & \quad\quad minus the known rotator J203349 & 7
\enddata
\tablecomments{Novelty means no previous compact-binary classification (Section~\ref{sec:comparison}). WD-CS denotes the white-dwarf cooling sequence; the DESI split follows Section~\ref{sec:desi_results}. The final 7-source high-priority subset is listed in Table~\ref{tab:high_priority}.}
\end{deluxetable}

\subsection{Hertzsprung--Russell Diagram}
\label{sec:hrd}

\begin{figure*}[htbp]
\centering
\includegraphics[width=0.66\textwidth]{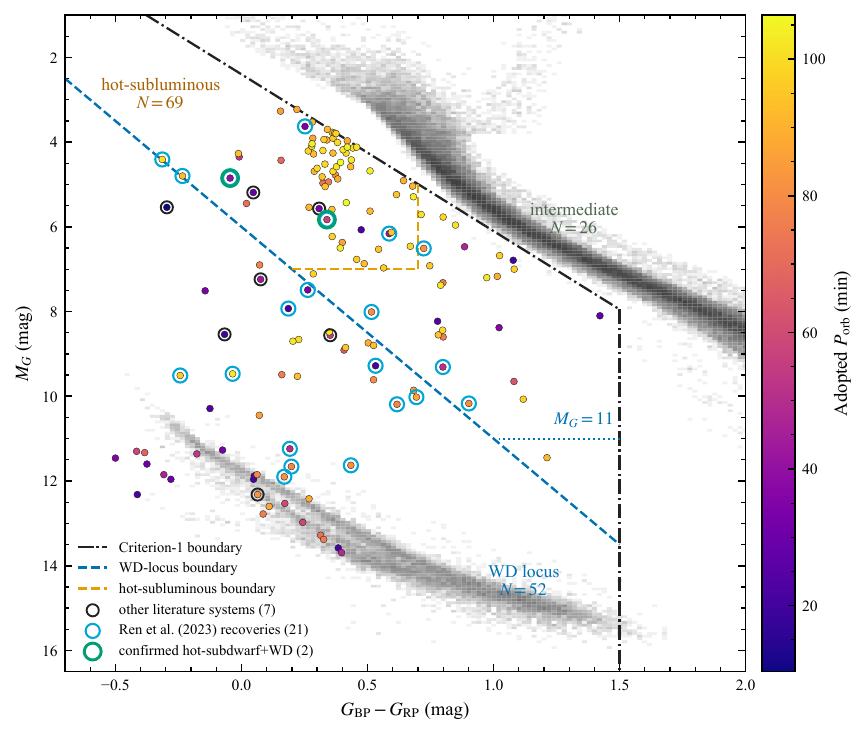}
\caption{Gaia color--magnitude diagram of all 147 catalog candidates, color-coded by the adopted $\Porb$, overlaid on the Gaia DR3 field-star density. The blue and orange dashed boundaries mark the white-dwarf locus and hot-subluminous region, respectively. Black dash--dotted lines show the initial below-main-sequence selection of Criterion~1, and the dotted line marks $\MG=11$. Open colored circles identify previously characterized compact systems. The region populations are listed in Table~\ref{tab:composition}.}
\label{fig:hr}
\end{figure*}

Figure~\ref{fig:hr} places all 147 candidates on the Gaia color--magnitude diagram, overlaid on the Gaia DR3 field-star density. We use the white-dwarf-locus boundary of \citet{GentileFusillo2021}, the hot-subluminous region of \citet{Geier2019} and \citet{Culpan2022}, and the original below-main-sequence boundary of Criterion~1; the resulting region populations are listed in Table~\ref{tab:composition}. The white-dwarf boundary is deliberately inclusive of overluminous white-dwarf binaries; the intermediate region can contain sdA stars, pre-extremely-low-mass (pre-ELM) white dwarfs, and composite systems \citep{Pelisoli2018}.

These regions are population diagnostics, not spectroscopic classifications. Absolute magnitudes of the distant, low-parallax-precision sources are correspondingly uncertain, whereas the nearby white-dwarf-locus subset has substantially better parallax precision; the parallax signal-to-noise ratio $\varpi/\sigma_\varpi$ is therefore included in the released catalog. Geometric distances from \citet{BailerJones2021} are used for the DESI subset. The adopted $\MG=11$ line highlights the faint cooling-sequence follow-up subset.

The novelty fraction is highest in the hot-subdwarf region, but the white-dwarf locus contributes the most compact new systems (Table~\ref{tab:composition}): of the 15 WD-locus sources with $\Porb<40$~min, 10 are newly identified, and 8 of these additionally satisfy $\MG\geq11$ and $\varpi/\sigma_\varpi\geq5$, so their cooling-sequence luminosities are secure against parallax uncertainty. One of the eight, ZTF~J203349.81+322901.10, is the double-faced white dwarf of \citet{Caiazzo2023}, whose 14.97-min signal is a published rotation period; we exclude it from the follow-up list, leaving the seven entries of Table~\ref{tab:high_priority}. All seven show EW-like double-wave morphology; five carry a published spectral type (two DA, one DB, one DC, one uncertain) and only J071816 has a DESI spectrum (Section~\ref{sec:J071816}), but none has a time-resolved radial-velocity series. They are the highest-priority new targets for time-resolved spectroscopy, and---if their photometric periods are orbital---the most compact of them, led by J213957, are the strongest unconfirmed GW candidates of Section~\ref{sec:gw}. Section~\ref{sec:harmonic_test} examines how far that condition can be trusted.

\begin{deluxetable}{lccccc}
\tablewidth{0pt}
\tabletypesize{\scriptsize}
\tablecaption{Highest-priority newly identified WD-locus candidates\label{tab:high_priority}}
\tablehead{
  \colhead{ZTF Name} & \colhead{$\Porb$} & \colhead{$\Delta P_{\rm cons}$} & \colhead{$G$} & \colhead{$\MG$} & \colhead{$\varpi/\sigma_\varpi$} \\
  \colhead{} & \colhead{(min)} & \colhead{(min)} & \colhead{(mag)} & \colhead{(mag)} & \colhead{}
}
\startdata
J213957.39$-$124550.08 & 21.257063 & 0.000033 & 16.41 & 13.58 & 461 \\
J071816.38$+$373138.66 & 22.545165 & 0.000037 & 16.95 & 12.32 & 124 \\
J071330.91$-$012623.33 & 22.560299 & 0.578705 & 16.84 & 11.96 & 158 \\
J164929.71$-$243310.22 & 24.024330 & 1.701743 & 17.15 & 11.88 & 107 \\
J215644.80$+$613633.37 & 29.917594 & 0.000065 & 18.99 & 11.60 & 21 \\
J045707.49$+$051322.03 & 32.499954 & 0.000080 & 18.31 & 11.27 & 26 \\
J192442.96$+$310403.62 & 35.869013 & 0.000093 & 19.37 & 11.46 & 11 \\

\enddata
\tablecomments{Newly identified WD-locus candidates with $\Porb<40$~min, $\MG\geq11$, and $\varpi/\sigma_\varpi\geq5$, ordered by $\Porb$. Eight sources meet these cuts; ZTF~J203349.81+322901.10 is omitted because its 14.97-min signal is the published rotation period of a double-faced white dwarf \citep{Caiazzo2023}. $\Porb$ is the adopted photometric orbital-period estimate and $\Delta P_{\rm cons}$ its conservative uncertainty (Section~\ref{sec:period_def}); the periods of J071330 and J164929 carry band-to-band uncertainties of 0.6--1.7~min and should be treated as approximate. \added{Periods are printed at the full precision needed to phase-fold the ZTF baseline: over $\sim$$10^{5}$ cycles a two-decimal period drifts the fold by tens of cycles.} All seven show EW-like double-wave morphology, and only J071816 has a DESI DR1 spectrum (Section~\ref{sec:J071816}). All entries require radial-velocity confirmation, and Section~\ref{sec:harmonic_test} shows that rotation is a viable alternative to an orbital origin for all of them. \added{These seven rows are an excerpt of the machine-readable catalog, which is published in its entirety and carries every source and column.}}
\end{deluxetable}

\begin{figure}[htbp]
\centering
\includegraphics[width=\columnwidth]{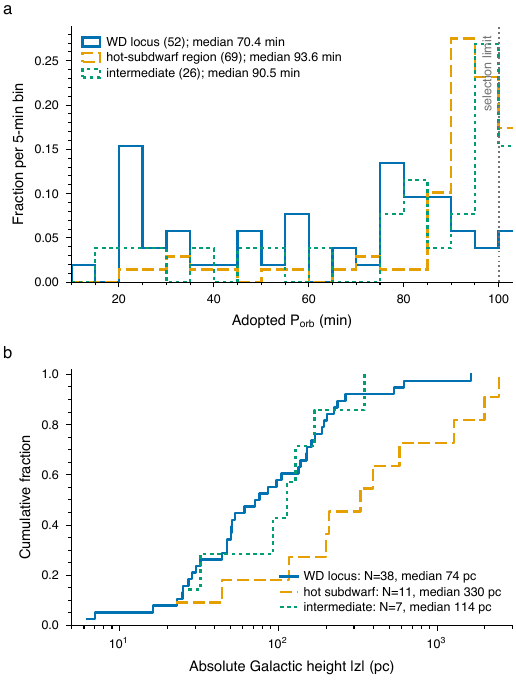}
\caption{Sample-level distributions by Gaia CMD region. \textit{(a)} Normalized distributions of the adopted photometric orbital-period estimates for the 147 candidates; the dotted line marks the 100-min search boundary on the repeating period $\Pref$, which the adopted $\Porb$ of double-wave EW-like systems can exceed (Section~\ref{sec:period_def}). \textit{(b)} Cumulative distributions of the absolute Galactic height $|z|=d\,|\sin b|$ for the 56 sources with $\varpi/\sigma_\varpi\geq5$, computed with $d=1000/\varpi$~pc. Panel (b) is a geometric consistency check, not a scale-height measurement (Section~\ref{sec:spatial_period}).}
\label{fig:period_distribution}
\end{figure}

Figure~\ref{fig:period_distribution}(a) shows the adopted-period distributions of the three CMD groups; panel~(b) shows their Galactic-height distributions, discussed in Section~\ref{sec:spatial_period}.

\subsection{DESI DR1 Spectroscopic Characterization}
\label{sec:desi_results}

The DESI subset provides an empirical view of the populations admitted by the photometric selection. Table~\ref{tab:desi} lists the 26 matches with adopted $\Porb$, Gaia photometry, and Bailer--Jones geometric distances \citep{BailerJones2021}. Gaia luminosities and DESI spectral morphology separate the subset into:
\begin{itemize}
  \item 9 sources on the white-dwarf cooling sequence ($\MG\gtrsim11$);
  \item 14 luminous-blue systems too bright for white dwarfs at their Gaia distances ($\MG\approx4$--$9$; hot-subdwarf/blue-horizontal-branch (BHB)/A-type candidates);
  \item 3 special or previously known systems (ES~Cet, 1RXS~J180804, ZTF~J110045.15+521043.71).
\end{itemize}
Thus only about one-third of the spectroscopically observed blue short-period variables lie on the white-dwarf cooling sequence. Section~\ref{sec:new_desi_sources} discusses the luminous-blue members, and Section~\ref{sec:individual} presents representative sources.

The single-epoch DESI spectra support qualitative line classification but not spectroscopic $\Teff$, $\logg$, or masses, which require higher-S/N follow-up. Appendix~\ref{app:panels} presents the full DESI spectra and ZTF phase-folded light-curve panels (Figures~\ref{fig:panelB1}--\ref{fig:panelB3}). Restricting confident line classifications to S/N$_R > 10$, the spectra fall into four morphological classes:

\begin{enumerate}
\item \textit{Broad Balmer-absorption spectra}: ZTF~J152934.91+292801.87 (S/N$_R = 55.8$; Figure~\ref{fig:panelB1}) and ZTF~J162009.42+125647.33 (S/N$_R = 18.7$) show broad H$\alpha$--H$\delta$ absorption. At low resolution this morphology is consistent with a hydrogen-rich photosphere, but a single DESI epoch alone does not establish a white-dwarf identification against luminous-blue alternatives, nor does it demonstrate a binary companion \citep{Pelisoli2018,Koester2010}. The absence of H$\alpha$ emission or He\,\textsc{ii} $\lambda4686$ argues against a high-accretion state in these epochs, although low-state accretion cannot be excluded.

\item \textit{Featureless continua} (9 sources): Sources such as ZTF~J071816.38+373138.66 (S/N$_R = 49.1$) and ZTF~J000637.94+310415.53 (S/N$_R = 15.5$) show no detectable spectral features across the full DESI wavelength range. Possible interpretations include: (a) a cool DC-type WD with $\Teff \lesssim 5000$~K where hydrogen recombination suppresses Balmer lines \citep{Bergeron1997}; (b) a helium-dominated DB/DC atmosphere at any temperature; or (c) a composite WD+companion system in which neither component produces strong features at optical wavelengths. Distinguishing these scenarios requires UV or near-IR photometry and higher-resolution spectroscopy.

\item \textit{Emission-line or accretion-like spectra}: ZTF~J160335.93+215032.33 shows narrow H$\alpha$ emission superposed on a blue continuum and broad Balmer absorption, consistent with an accreting or irradiated eclipsing WD binary candidate. ZTF~J180805.18+581011.72 (1RXS~J180804.3$+$581001) shows emission-like Balmer morphology and is retained as an X-ray/CV-like system rather than a detached-DWD candidate.

\item \textit{Composite or uncertain} (low S/N): Sources with S/N$_R < 10$ including ZTF~J000208.56+093543.14 cannot be reliably classified from the available spectra. We flag these as requiring deeper follow-up.
\end{enumerate}

\subsection{Light Curve Morphology and Sample Classification}
\label{sec:lc_class}

The ranking labels follow the standard light-curve terminology. EA-like curves show relatively narrow minima and a flatter inter-eclipse baseline, whereas EW-like curves show continuous, approximately double-wave modulation. These are descriptive morphology labels, not physical claims about contact configuration. Among the 26 DESI sources, 22 are EW-like and 4 are EA-like. ES~Cet has the shortest period ($\Porb=10.34$~min; \citealt{Espaillat2005}); among the unconfirmed DESI members, J071816 has the shortest adopted estimate ($\Porb=22.55$~min).

\subsection{Notes on Individual Objects}
\label{sec:individual}

\begin{figure*}[htbp]
\centering
\includegraphics[width=0.9\textwidth]{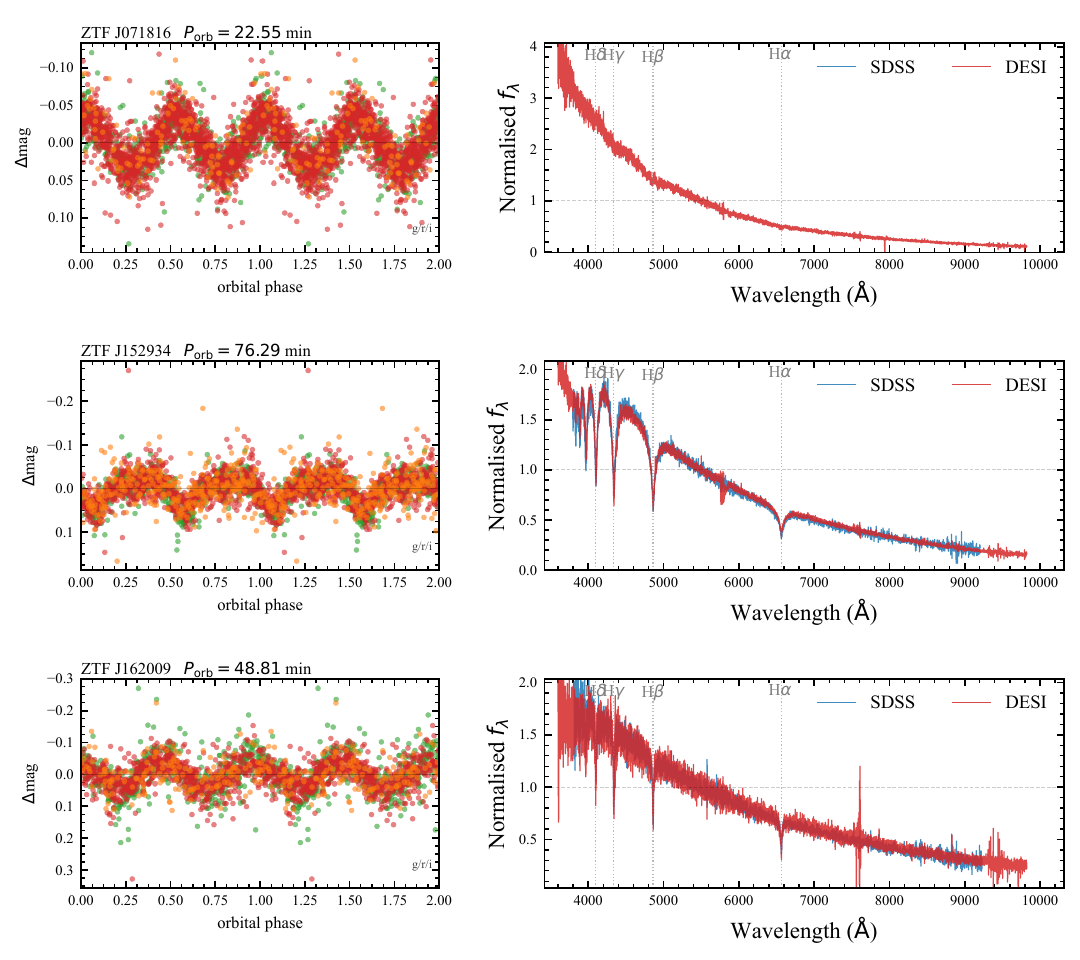}
\caption{Compact panels for the three non-eclipsing discussed sources, one per row: J071816, J152934, and J162009 (top to bottom). \textit{Left}: ZTF light curves folded at the adopted $\Porb$ ($\Delta$ magnitudes relative to each band's median; two cycles shown). \textit{Right}: normalized DESI spectra with Balmer lines marked; archival SDSS spectra are overlaid where available. The deep eclipser J160335 is shown separately in Figure~\ref{fig:J160335}.}
\label{fig:featured_trio}
\end{figure*}

\begin{figure*}[htbp]
\centering
\includegraphics[width=0.85\textwidth]{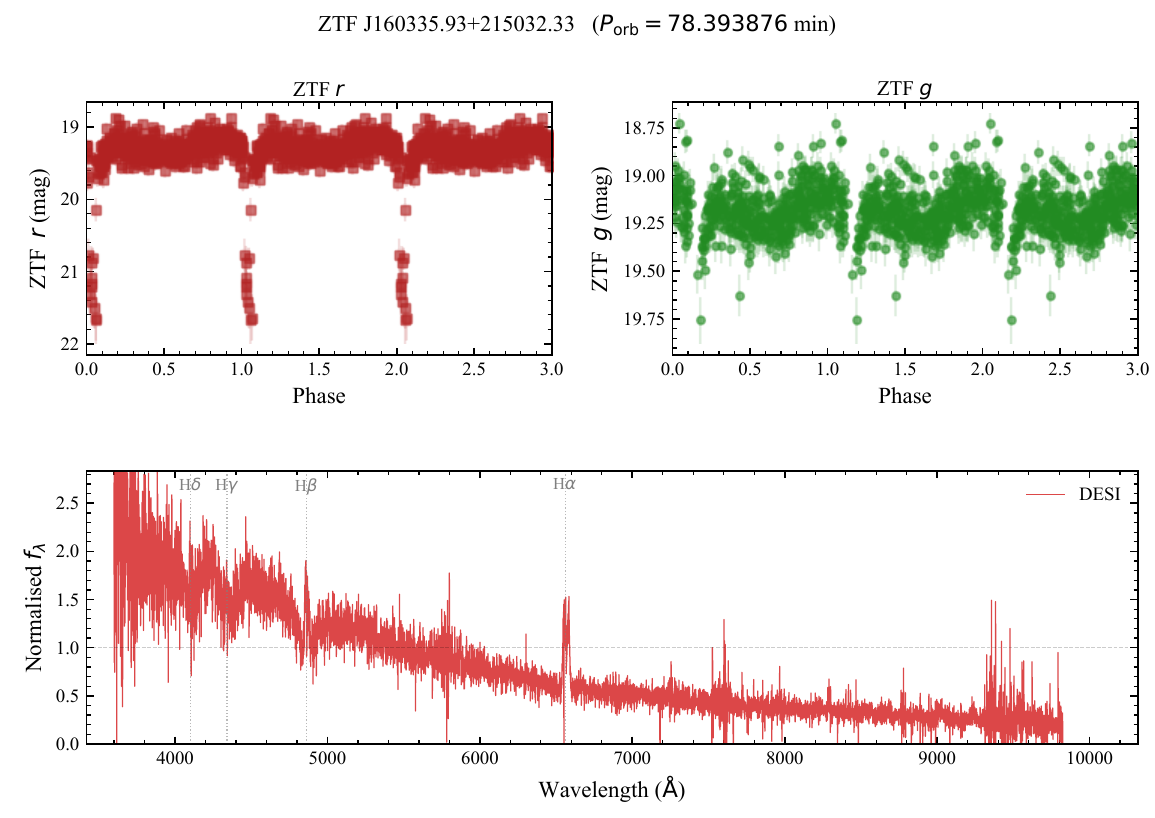}
\caption{The deep-eclipsing system ZTF~J160335.93+215032.33. \textit{Top}: phase-folded ZTF $r$- and $g$-band light curves at the adopted $\Porb=78.39$~min; the catalog photometry reaches at least 2.3~mag below the $r$-band baseline. \textit{Bottom}: normalized DESI spectrum, showing a blue continuum with H$\alpha$ emission.}
\label{fig:J160335}
\end{figure*}

We use four WD-locus DESI members---J071816, J160335, J152934, and J162009---to illustrate the range of photometric and spectral properties. J160335 also appears in \citet{Ren2023}, while the other three had only photometric white-dwarf catalog entries \citep{GentileFusillo2021}. The deep eclipser J160335 is shown individually in Figure~\ref{fig:J160335}; the three non-eclipsing sources share the compact panels of Figure~\ref{fig:featured_trio}. All panels use the source-specific, individually refined, full-precision $\Porb$ estimates defined in Section~\ref{sec:period_def}. For the non-eclipsing sources, we quantify the modulation as the 5th--95th percentile range of the medians in 30 equal phase bins, measured separately in $g$ and $r$. As a long-term stability check we also split each light curve at its median epoch and compute the Pearson correlation $r_{g}$, $r_{r}$ between the phase-binned profiles of the two halves, each spanning roughly three years.

\subsubsection{ZTF J071816: a UV-bright white-dwarf candidate}
\label{sec:J071816}

J071816 lies on the Gaia white-dwarf sequence at $d=85$~pc and $\MG=12.32$ (Figure~\ref{fig:featured_trio}, top row). Folding at $\Porb=22.55$~min produces a coherent double-wave modulation, with phase-binned amplitudes of approximately 0.065~mag in $g$ and 0.068~mag in $r$, and the phase profile is stable across the two halves of the 6.6-yr baseline ($r_g=0.94$, $r_r=0.95$). The nearly achromatic, long-lived modulation disfavors a strongly temperature-dependent process such as a dominant reflection effect. It cannot, however, be ellipsoidal: \citet{GentileFusillo2021} fit the source at $M=1.33\,\Msun$ and $\logg=9.26$, so its radius is $0.0045\,R_\odot$ and the maximum tidal amplitude at a 22.5-min orbit is $3\times10^{-6}$~mag, four orders of magnitude below what is observed (Section~\ref{sec:harmonic_test}). A stable surface structure carried around by rotation is the natural remaining explanation. The DESI spectrum is featureless from B to Z at S/N$_R=49.1$, with no detectable Balmer, He\,\textsc{i}, Ca\,\textsc{ii}, or emission lines---consistent with the published DC classification---while the GALEX photometry shows that the source is markedly ultraviolet bright, with ${\rm FUV}-{\rm NUV}=-0.329\pm0.019$ and ${\rm FUV}-G=-1.568\pm0.018$; these colors require a hot photospheric component and exclude a cool-DC interpretation. A hot, ultramassive, featureless white dwarf with a coherent 11.27-min photometric signal closely resembles ZTF~J1901+1458 \citep{Caiazzo2021}, and we regard a rapidly rotating magnetic white dwarf as the leading interpretation. Time-resolved spectropolarimetry would settle it; if the modulation is instead orbital, J071816 has the highest conditional GW ranking among the unconfirmed WD-locus candidates (Section~\ref{sec:gw}).

\subsubsection{ZTF J160335: a deep-eclipsing candidate with H$\alpha$ emission}
\label{sec:J160335}

J160335 is the strongest eclipsing system among the newly examined WD-locus sources ($\MG\simeq11.7$, $d=308$~pc; Figure~\ref{fig:J160335}). At $\Porb=78.39$~min, 12 ZTF $r$-band measurements lying at least 0.5~mag below the out-of-eclipse median---obtained on 12 separate nights spanning 5.2~yr---are confined to $\Delta\phi=0.049$, corresponding to an observed eclipse window of approximately 3.8~min. Their depths range from 0.86 to 2.38~mag, individually 5.2--9.7 times the per-point photometric uncertainty, so the phase concentration of independent nightly detections across five years establishes a genuine recurrent eclipse rather than a sequence of outliers. The faintest catalog measurement is 2.38~mag below the baseline, implying that at least 89\% of the $r$-band flux is removed. Because the ZTF sampling does not resolve the eclipse bottom or the contact points, the depth is a lower limit and the phase interval is not a geometric contact duration. The DESI spectrum shows Balmer emission at two lines: local-continuum measurements give emission equivalent widths of approximately 14.7~\AA\ at 4.3$\sigma$ for H$\alpha$ and 7.3~\AA\ at 5.4$\sigma$ for H$\beta$. The recurrent deep eclipse and two-line Balmer emission favor an accreting or strongly irradiated WD binary rather than a clean detached-DWD interpretation. High-speed multiband eclipse photometry and phase-resolved spectroscopy can determine the geometry, locate the emission-line source, and establish a precise timing ephemeris.

\subsubsection{ZTF J152934: a nearby WD binary candidate}
\label{sec:J152934}

J152934 lies at $d=87$~pc with $\MG=12.79$ (Figure~\ref{fig:featured_trio}, middle row). Its $\Porb=76.29$~min phase curve has binned amplitudes of approximately 0.079~mag in $g$ and 0.065~mag in $r$, stable across the two halves of the baseline ($r_g=0.93$, $r_r=0.89$); the mild blue-band enhancement is still close to the achromatic expectation of geometric modulation. Both the archival SDSS spectrum and the DESI spectrum show broad H$\alpha$--H$\delta$ absorption with closely similar morphology, and neither epoch shows significant H$\alpha$ or He\,\textsc{ii} emission, arguing against a high-accretion state. The same SDSS spectrum, however, already yielded a magnetic DAH classification \citep[CSO~1094;][]{Kleinman2013}, and at $M=0.96\,\Msun$ the largest ellipsoidal amplitude a 76.3-min orbit could produce is $3\times10^{-6}$~mag against the observed 0.079~mag. The double-wave light curve is therefore far more naturally the rotation of a magnetic white dwarf than a detached binary, and we reclassify it as such pending phase-resolved spectropolarimetry. It remains above the adopted fiducial LISA threshold only under the orbital hypothesis.

\subsubsection{ZTF J162009: the nearest DESI-matched candidate}
\label{sec:J162009}

At $d=75$~pc and $\MG=13.69$, J162009 is the nearest DESI-matched WD-locus candidate (Figure~\ref{fig:featured_trio}, bottom row). Folding at $\Porb=48.81$~min gives phase-binned amplitudes of approximately 0.133~mag in $g$ and 0.084~mag in $r$, and the half-baseline profile correlations ($r_g=0.82$, $r_r=0.84$) are lower than for J071816 and J152934. The stronger blue-band modulation, roughly 1.6 times larger in $g$ than in $r$, distinguishes it from the nearly achromatic variability of J071816 and makes temperature-dependent mechanisms, including pulsation, spots, or irradiation, competitive alternatives to purely geometric modulation. Both SDSS and DESI spectra show Balmer absorption without significant emission, supporting a hydrogen-atmosphere white dwarf but not establishing orbital motion. Its proximity makes it an efficient target for simultaneous high-cadence multiband photometry and time-resolved spectroscopy.

The four cases span the range the selection admits: J160335 is a strong eclipsing accreting or irradiated WD binary, while for J071816, J152934, and J162009 rotation of a magnetic or chemically inhomogeneous white dwarf competes with, and in the first two cases is preferred over, a binary origin (Section~\ref{sec:harmonic_test}).

\subsubsection{Recovered known systems}
\label{sec:recovered}
\label{sec:J020052}
\label{sec:J1100}

Five previously studied systems are recovered by the pipeline and serve as validation. ES~Cet (ZTF~J020052.25$-$092431.69) is an established AM~CVn verification binary; the refined ZTF period, $10.336880\pm0.000016$~min ($620.213$~s), agrees with its published 620.2-s orbit \citep{Espaillat2005,Bakowska2021,Kupfer2018} and stays phase-coherent across the full ZTF baseline. ZTF~J110045.15+521043.71 is a \citet{Ren2023} compact binary whose DESI spectrum shows a five-line Balmer series. ZTF~J103533.01+055159.00 and ZTF~J214140.42+050729.92 are cataloged cataclysmic variables \citep{Inight2023} recovered as deep eclipsers at 82.09 and 78.76~min. ZTF~J203349.81+322901.10 is the double-faced white dwarf of \citet{Caiazzo2023}: the pipeline recovers its 14.97-min signal, which is the published \emph{rotation} period of a single magnetic star rather than an orbit, and it is excluded from the follow-up and gravitational-wave rankings for that reason. Its recovery is the clearest in-sample demonstration that a coherent minute-scale double-wave fold does not by itself imply a binary (Section~\ref{sec:harmonic_test}). Among the photometric-only members, the pipeline also independently recovers the two shortest-period systems after ES~Cet: the 20.5-min ultracompact binary TMTS/ZTF~J0526+5934 \citep{Lin2024,Rebassa2024a} and the 20.6-min detached binary PTF~J0533+0209 \citep{Burdge2019b}.

\subsubsection{Luminous-blue members: hot subdwarfs, BHB, and A-type stars}
\label{sec:new_desi_sources}

The 14 luminous-blue members illustrate the heterogeneous population selected by these criteria (Section~\ref{sec:selection_function}). Eight DESI sources show a four-line Balmer series at $\geq5\%$ continuum depth, yet seven of them have $\MG\approx4$--$7.5$ at $d=1.9$--$6.7$~kpc (Table~\ref{tab:desi})---4--9~mag above the WD cooling sequence. A WD interpretation would require the Gaia parallaxes to be in error by factors of 10--25, which is excluded. Their $\MG$, colors, and Balmer absorption are instead consistent with hot subdwarfs (sdB/sdO), blue-horizontal-branch, or A-type stars, whose low-resolution spectra can mimic broad Balmer-absorption white-dwarf morphologies \citep{Pelisoli2018}; Gaia $\MG$ is the discriminant. Panels appear in Figures~\ref{fig:panelB2}--\ref{fig:panelB3}.

The clearest example is ZTF~J213611.54+092159.85 ($G=16.88$, S/N$_R=30.8$, $d=4.2$~kpc, $\Porb=85.22$~min): its $\MG=3.75$ places it firmly among the hot subdwarfs. Short-period sdB/sdO binaries such as the sdB+WD systems ZTF~J2130+4420 and ZTF~J2055+4651 \citep{Kupfer2020} share this photometric signature, so a blue, short-period, Balmer-lined source can be a hot-subdwarf binary rather than a white-dwarf binary.

\section{Discussion}
\label{sec:discussion}

\subsection{Sample Composition, Selection Effects, and Follow-up}
\label{sec:completeness}
\label{sec:two_samples}
\label{sec:photometric_limitations}

The Gaia criteria recover $209\rightarrow187\rightarrow169\rightarrow166\rightarrow159\rightarrow122$ of the reference set, a final retention of 122/209 (58.4\%), with the largest single loss (37 sources) at the variability cut, where intrinsically variable but faint systems fall below the Gaia photon-noise floor.

\added{Table~\ref{tab:completeness} traces where the 122 Gaia survivors are lost. Thirty of the 209 reference positions have no ZTF DR23 counterpart (28 of them south of the survey boundary at $\delta=-28^\circ$), and five more have fewer than the 50 usable epochs the permutation test requires; the footprint and cadence are independent of the Gaia criteria, so these fractions remove about 18 and 2 of the 122, leaving of order 102 that reach the period search. The search window removes none: all 209 published reference periods lie between 5.4 and 99.4~min, inside the 3--100~min window. The remaining loss, from about 102 to the 28 recovered, is the combined cost of detection significance, CNN ranking, and the refolding vetting, and is the largest single term after the Gaia cuts.}

\added{To estimate how many of these signals are detectable in ZTF DR23 at all, we re-ran the detection-stage multi-harmonic AoV periodogram on the 169 sufficiently sampled reference systems (the 174 of Table~\ref{tab:completeness} minus duplicate literature entries and one baseline shorter than 100~d). The published signal, or its first harmonic or subharmonic, is the strongest feature in the 3--100~min band for 24 systems (14\%) and exceeds the 99.99th percentile of the periodogram background for 50 (30\%). Scaled to the $\sim$102 systems reaching the search, of order 30 carry a detectable signal, comparable to the 28 recovered, so the depth of the data is a leading term in the residual. The overlap is imperfect---19 of the 27 recovered systems in the test sample are among the 50 flagged, and 31 flagged systems were not recovered---so ranking and vetting losses remain entangled with detectability. The non-recovered systems are fainter (median $G=18.30$ against 17.97), more sparsely sampled (542 against 882 usable epochs), and longer-period (82.0 against 56.3~min) than the recovered ones. The test is single-band and misses 8 of the 27 recovered systems it covers; separating the detection, ranking, and vetting terms cleanly would require an injection--recovery experiment, which we have not attempted.}

\begin{deluxetable}{lccll}
\tablewidth{0pt}
\tabletypesize{\scriptsize}
\tablecaption{\rev{Completeness budget for the 209-source reference set}\label{tab:completeness}}
\tablehead{
  \colhead{Stage} & \colhead{$N$} & \colhead{of 209} & \colhead{Lost} & \colhead{Cause}
}
\startdata
Reference set & 209 & 100.0\% &  &  \\
Gaia five criteria & 122 & 58.4\% & 87 & Table 1; dominated by the variability cut \\
ZTF DR23 counterpart & 104 & 49.8\% & 18 & outside the ZTF footprint \\
$\geq$50 epochs in $g$ or $r$ & 102 & 48.8\% & 2 & below the minimum sampling \\
Period inside the search window & 102 & 48.8\% & 0 & no loss: all reference periods are inside 3--100 min \\
Recovered end to end & 28 & 13.4\% & 74 & period search, CNN ranking, and refinement \\

\enddata
\tablecomments{\rev{The Gaia stage is the published chain of Table~\ref{tab:gaia_criteria}, evaluated against the Gaia archive at selection time. It is not re-derived from the reference-set table released here, in which six sources lack a positive parallax and seven literature systems enter through two Gaia matches each. The two ZTF terms are measured by querying ZTF DR23 at all 209 reference positions---179 have a counterpart and 174 have at least 50 usable epochs---and applying those fractions to the 122 Gaia survivors, which is justified because the ZTF footprint and cadence are uncorrelated with the Gaia criteria. The search window removes nothing: all 209 reference periods lie between 5.4 and 99.4~min. The final line therefore isolates the combined detection, ranking, and vetting loss. The per-source stage flags are released with the data products.}}
\end{deluxetable}

The catalog is neither volume limited nor complete for any physical binary class: it requires a blue subluminous Gaia position, acceptable astrometry and photometry, excess Gaia variability, sufficient ZTF coverage, and a coherent 10--100~min modulation. The 122/209 recovery fraction therefore measures the behavior of the pipeline on its heterogeneous reference set, not a Galactic binary fraction, formation rate, or merger rate; the latter would require forward modeling of the survey and variability selection functions \citep{Burdge2020a,Ren2023}. The selection favors high-inclination eclipsing systems, tidally distorted stars, and reflection- or accretion-driven variables, and disfavors nearly face-on detached binaries, weakly modulated systems, and objects dominated by a red companion. The final catalog is correspondingly broader than a double-white-dwarf sample: the same short-period photometric signature is shared across the WD-locus, hot-subdwarf-region, and intermediate populations (Table~\ref{tab:composition}), so we use \emph{compact-binary candidate} for the full catalog and reserve \emph{confirmed orbit} \added{for the three recovered systems with independent time-resolved evidence---ES~Cet, TMTS/ZTF~J0526+5934, and PTF~J0533+0209}.

The 26 DESI spectra are likewise not a random subsample, since DESI targeting favors blue stars. Their 9 white-dwarf-cooling-sequence, 14 luminous-blue, and 3 known or peculiar systems nonetheless expose the central degeneracy of the parent selection: similar short-period light curves arise in DWD/AM~CVn candidates, accreting white-dwarf binaries, and hot-subdwarf binaries. A Balmer-absorption spectrum does not by itself break this degeneracy, and a featureless or single-epoch spectrum does not exclude a compact binary. The four spectral morphologies of Section~\ref{sec:desi_results} therefore match the mix expected from the selection function rather than measuring intrinsic class fractions, and the absence of H$\alpha$ or He\,\textsc{ii} $\lambda4686$ emission in most epochs argues against high-state accretion without excluding low-state systems.

The catalog nevertheless provides a practical confirmation sequence. The 52 WD-locus candidates, particularly the 15 with $\Porb<40$~min, are the most direct targets for time-resolved spectroscopy and high-cadence multiband photometry; within these, the seven newly identified sources with secure cooling-sequence luminosities listed in Table~\ref{tab:high_priority}---five of which already carry a published spectral type, but none of which has a time-resolved radial-velocity series---are the natural starting point. Appendix~\ref{app:lconly} shows the 18 non-DESI sources with $\MG\geq11$ selected for immediate follow-up. Deep eclipsers can yield component radii and inclinations when eclipse modeling is combined with radial velocities, while coherent non-eclipsing systems require an RV curve to establish whether the modulation traces orbital motion. For luminous-blue sources, phase-resolved spectroscopy can separate sdB+WD systems from sdB+dM or pulsating-star alternatives through their velocity amplitudes, line profiles, and wavelength-dependent variability. The remaining 121 candidates without DESI spectra are the principal discovery and follow-up pool rather than a confirmed-binary sample.

\subsection{Relation to Previous Searches and Scientific Use of the Catalog}
\label{sec:comparison}

Of the 23 \citet{Ren2023} candidates with $\Porb<100$~min, we recover 21 (91\%). The remaining two are lost during period recovery or Gaia photometric-quality selection. Because the Ren et al.\ sample contributes to the calibration context, this is a recovery check rather than a blind validation.

Across all literature and catalog cross-matches, 36 of the 147 sources have a previous compact-binary classification and 111 (75\%) are newly identified compact-binary candidates. Of those 111, 27 appear in a general variable-star catalog, usually without a compact-object classification, and 84 have no previous variable-star entry. Thus ``new'' consistently means no prior compact-binary classification.

The closest photometric comparison is \citet{Ren2023}, who used Gaia EDR3 and ZTF DR8 to assemble 429 candidates across several close-white-dwarf-binary morphologies and a wider period range. Our catalog deliberately concentrates on the 10--100~min regime using Gaia DR3 and the longer ZTF DR23 baseline. It is therefore smaller but more focused on the shortest-period population: 52 objects occupy the WD locus, including 15 with $\Porb<40$~min, while 69 occupy the hot-subdwarf region, including five with $\Porb<40$~min. These counts are observed-sample properties, not intrinsic relative occurrence rates. Relative to earlier Gaia/ZTF searches, the principal methodological differences are:

\begin{enumerate}
\item Gaia thresholds are calibrated against known SPWDBs with a retention-rate ratio (Appendix~\ref{app:gaia_criteria}).
\item The CNN ranks EA/EW-like folded ZTF light curves (94.3\% held-out validation accuracy), after which every promoted target is refolded at a high-precision refined period.
\item DESI spectra for 26 members quantify the WD versus luminous-blue mix admitted by the photometric cuts (Section~\ref{sec:desi_results}).
\end{enumerate}

Recent spectroscopic searches provide a complementary selection. \citet{Jiang2025} used DESI multi-epoch radial velocities to identify DA DWD candidates, and \citet{Pallathadka2025} used SDSS-V sub-exposure velocities. Their approach supplies direct evidence for orbital motion and supports atmospheric and binary-population inference, but it is naturally strongest for line-bearing DA white dwarfs with adequate spectral S/N and repeated observations. The present photometric route is most sensitive to favorable inclinations and detectable surface-brightness modulation, but can select eclipsers, non-DA or optically featureless WD-locus objects, and targets outside existing multi-epoch spectroscopic footprints. The two approaches thus probe different observational projections of the same short-period population rather than competing definitions of a DWD.

The scientific use of the catalog follows its CMD substructure. Confirmed short-period WD-locus binaries would enlarge the sample available for common-envelope, mass-transfer, eclipse-timing, and mHz-GW studies. Confirmed hot-subdwarf binaries would instead constrain post-common-envelope sdB+WD and sdB+dM channels; the shortest-period members are especially relevant because systems such as ZTF~J2130+4420 and ZTF~J2055+4651 demonstrate that this photometric region contains compact sdB+WD binaries \citep{Kupfer2020}. The 26 intermediate objects are natural targets for identifying pre-ELM, composite, and transition systems \citep{Pelisoli2018}. Simple ultraviolet and mid-infrared color diagnostics for the 74-source SED subsample---most of it without DESI coverage---identify hot photospheric components and flag two short-period candidates with significant $W1-W2$ excesses (Appendix~\ref{app:uvir}). The catalog thus provides a structured list of 111 new candidates spanning several compact-binary evolutionary channels; for WD-locus sources without H$\alpha$/He~II emission, a detached DWD interpretation remains viable but unconfirmed.

\subsection{The Photometric Harmonic and the Rotating White-Dwarf Alternative}
\label{sec:harmonic_test}

Adopting $\Porb=2\Pref$ for a double-wave source assumes the modulation is ellipsoidal. Two independent checks show that assumption is not supported by the light curves themselves for most of the catalog, and that a rotating, chemically or magnetically inhomogeneous single white dwarf is a competitive---in several cases preferred---interpretation.

\textit{Is the double wave real?} A tidally distorted binary has two physically distinct minima, so its Fourier spectrum contains the fundamental $f_1=1/\Porb$ as well as the harmonic $f_2=1/\Pref$. A single-mode pulsation or a spotted rotator produces power only at $f_2$; folding it at $2\Pref$ manufactures two identical humps. For the 69 catalog sources with $\Porb=2\Pref$ and archival $g$ and $r$ coverage we measure $R=P(f_1)/P(f_2)$ from the joint periodogram and the difference between the two halves of the $\Porb$ fold. Fifty-six have $R<0.05$ and 54 have half-cycle profiles agreeing within $2\sigma$: for these the data carry no positive evidence that the repeating period is half an orbit. Only eight satisfy both $R>0.15$ and a $>3\sigma$ half-cycle difference, and that set is dominated by the independently confirmed systems---the \citet{Ren2023} eclipsing binary ZTF~J110045.15+521043.71 ($R=34.6$) and the \citet{Kupfer2020} sdB+WD binaries ZTF~J2130+4420 and ZTF~J2055+4651. The test is therefore a useful \emph{promotion} criterion. Two of the eight are not previously classified: J232029 and J035315, the latter matching the ROSAT source 1RXS~J035315.5+095700, which for the same reason we treat as an X-ray/CV-like system rather than a detached-DWD candidate, as we do for 1RXS~J180804. The test is not a rejection criterion: a detached binary in which only one component is distorted also produces a nearly pure harmonic, and the confirmed systems TMTS/ZTF~J0526+5934 and PTF~J0533+0209 both fail it.

\textit{Can the amplitude be ellipsoidal at all?} This is the stronger constraint. Ellipsoidal amplitude scales as $(R_1/a)^3$ \citep{Morris1985}, and $R_1$ follows from the published photometric radius. For the 28 WD-locus members with pure-hydrogen fits in \citet{GentileFusillo2021} we compute the maximum peak-to-peak ellipsoidal amplitude at the adopted $\Porb$, generously assuming an equal-mass companion and an edge-on orbit. The median is $1.2\times10^{-5}$~mag, and only two sources exceed $0.01$~mag: PTF~J0533+0209 and J211119, the two extremely low-mass ($M\approx0.15\,\Msun$, $R\approx0.05\,R_\odot$) members---and PTF~J0533+0209 is precisely the one confirmed ellipsoidal double white dwarf in the sample. For the compact, high-gravity majority, including the featured sources J071816, J152934, J162009, and J213957, the observed 0.05--0.13~mag modulations exceed any possible tidal signal by three to five orders of magnitude. Tidal distortion of a white dwarf of this radius simply cannot produce them.

\textit{What can?} A surface brightness inhomogeneity carried around by rotation reproduces every observed property: a single wave at the rotation period, no fundamental at twice that period, an amplitude set by spot contrast rather than by $(R_1/a)^3$, stability over years, and a mild blue enhancement. Rapidly rotating magnetic and double-faced white dwarfs are an established class \citep{Ferrario2015,Caiazzo2021,Caiazzo2023}, and the catalog contains proven examples: ZTF~J203349.81+322901.10 is the double-faced white dwarf of \citet{Caiazzo2023}, whose $\Pref=14.97$~min is a published \emph{rotation} period, and J152934 carries a published magnetic DAH classification \citep{Kleinman2013}. Among the WD-locus members with published atmospheric parameters, \added{13} of 28 have $M\geq0.8\,\Msun$ and five are ultramassive ($M\geq1.0\,\Msun$), a marked excess over the field mass function and characteristic of the merger remnants that dominate the fast-rotating magnetic population; J071816 itself is fitted at $M=1.33\,\Msun$ and $\Teff\approx3.0\times10^{4}$~K, close to ZTF~J1901+1458 \citep{Caiazzo2021}. Non-radial pulsation is a further alternative where the star lies in an instability strip \citep{Corsico2019}, though most WD-locus members fall outside the DAV and DBV strips.

\begin{figure}[htbp]
\centering
\includegraphics[width=\columnwidth]{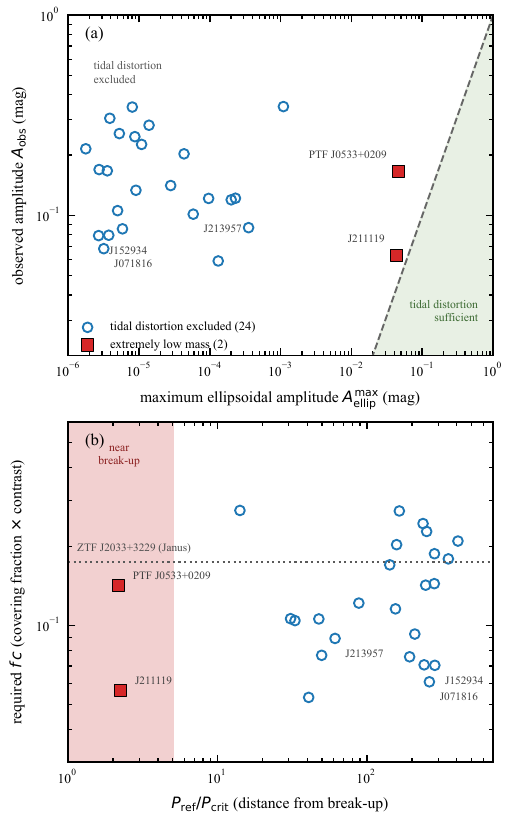}
\caption{\rev{The two interpretations tested on the same 26 WD-locus sources. \textit{(a)} Observed peak-to-peak ZTF amplitude against the maximum ellipsoidal amplitude permitted by the published radius at the adopted $\Porb$; the dashed line is equality and the shaded wedge is where tidal distortion suffices. Only the two extremely low-mass members approach it, the remaining 24 falling short by 2.4--5.1 orders of magnitude. \textit{(b)} The same sources in the plane of the surface inhomogeneity the rotation hypothesis requires, $fc=1-10^{-0.4A_{\rm obs}}$, against how far the implied rotation lies from break-up. The dotted line marks the value required by ZTF~J203349.81+322901.10 (Janus), the confirmed double-faced rotator recovered by this pipeline and measured with the same estimator; the shaded band marks near-critical rotation, which only the two extremely low-mass members would need.}}
\label{fig:rotation}
\end{figure}

\added{We applied the same test to the rotation hypothesis, on the same 28 WD-locus members with published pure-hydrogen fits and with the amplitudes of Section~\ref{sec:individual}. Two sources whose folded profiles do not reproduce between the two halves of the ZTF baseline (Pearson $r\leq0.3$ in at least one band) are excluded, leaving 26. Table~\ref{tab:rotation} lists the per-source values and Figure~\ref{fig:rotation} shows the two hypotheses together.}

\added{Setting $P_{\rm rot}=\Pref$, the 24 compact members require $v_{\rm eq}=2\pi R_1/\Pref=9$--$128$~km~s$^{-1}$ and $\Pref/P_{\rm crit}=14$--$407$ (median 179), where $P_{\rm crit}=2\pi(R_1^{3}/GM_1)^{1/2}$ is the Keplerian break-up period: two orders of magnitude slower than break-up. Minute-scale rotation is fast for a white dwarf but dynamically unproblematic, and is realised in the fast rotators above. A peak-to-peak amplitude $A$ demands a projected covering fraction times band contrast $fc=1-10^{-0.4A}=0.053$--$0.275$ (median 0.119), i.e.\ $\Delta T/T\approx10\%$ for a hemispheric feature at the published $\Teff$. The catalog sets this scale internally: ZTF~J203349.81+322901.10, the double-faced white dwarf of \citet{Caiazzo2023}, requires $fc=0.17$ on the same estimator, and the magnetic DAH J152934 requires $fc=0.07$. The ellipsoidal hypothesis instead falls short in amplitude by $2.4$--$5.1$~dex (median 4.3); since $A_{\rm ellip}\propto(R_1/a)^3$, closing that gap at fixed separation would take radii larger than the measured ones by factors of $6$--$50$. The two exceptions are the extremely low-mass members PTF~J0533+0209 and J211119, at $\Pref=2.2\,P_{\rm crit}$; no white dwarf is known to rotate that close to break-up, and their radii are the only ones allowing an ellipsoidal amplitude within a factor $1.5$--$3.6$ of the observed one. The two hypotheses fail in complementary parts of the sample.}

\added{At the population level the amplitudes do not track $A_{\rm ellip}^{\rm max}$, which spans five decades here (Spearman $\rho=-0.08$, $p=0.70$), as expected when contrast rather than $(R_1/a)^3$ sets the amplitude. A control sample from the same catalog \citep{GentileFusillo2021}, matched to each source within 0.5~mag in $G$ and 0.05~dex in $\log\Teff$, predicts 3.8 members at $M\geq0.8\,\Msun$ and 0.9 ultramassive against the observed 13 and 5 ($p=2\times10^{-5}$ and $1\times10^{-3}$); it is not matched in the full selection function, so the excess is supporting rather than decisive. ZTF colors, finally, do not discriminate: a single-temperature feature predicts $A_g/A_r=1.09$--$1.27$ and geometric modulation predicts unity, while the measured ratios (median 1.07; 20 of 26 consistent with unity at $3\sigma$, four bluer, two---both accreting or eclipsing---redder) are compatible with both. Time-resolved spectropolarimetry and high-speed multiband photometry, not more survey photometry, are the decisive follow-up.}

We therefore keep $\Porb=2\Pref$ as the adopted binary-hypothesis value, because it is the correct reading if the modulation is ellipsoidal, but the catalog is best read as a sample of coherently variable blue compact objects in which a rotating single white dwarf and a compact binary are not separated by ZTF photometry alone. The harmonic ratio $R$, the half-cycle difference, and the maximum ellipsoidal amplitude are released as catalog columns so that this distinction can be applied directly. The fiducial gravitational-wave quantities of Section~\ref{sec:gw} inherit this caveat in full.

\subsection{Fiducial GW Estimates for TianQin, LISA, Taiji, and DECIGO}
\label{sec:gw}

We estimate the expected GW signals under two fiducial assumptions: the adopted photometric period is orbital, and the chirp mass is $\Mc=0.3\,\Msun$. We use the catalog $\Porb$ values directly and set $\fGW=2/\Porb$. \added{Only three catalog members---ES~Cet, TMTS/ZTF~J0526+5934, and PTF~J0533+0209---have independently established orbits, and none of the newly identified candidates does;} these estimates are intended only to prioritize follow-up, not to establish detectability. \added{The signal model, the adopted sensitivity curves, and the signal-to-noise definition follow \citet{Yu2026} and are collected in Appendix~\ref{app:gwmodel}.}

\begin{figure*}[ht]
  \centering
  \includegraphics[width=0.9\textwidth]{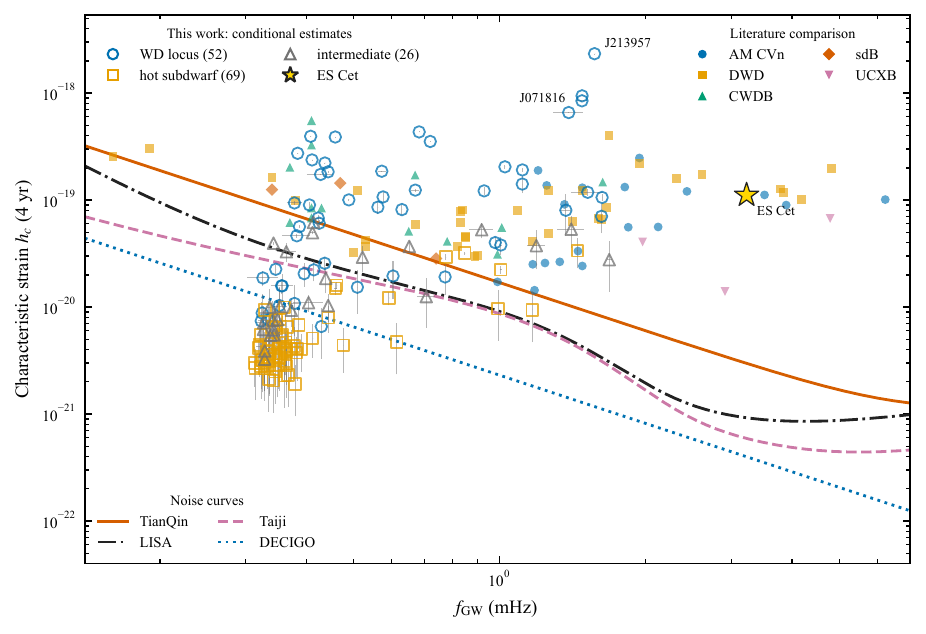}
  \caption{Fiducial characteristic strain versus GW frequency. Open circles show the 147 candidates using the adopted catalog $\Porb$, $\Mc=0.3\,\Msun$, and a 4-yr sky-averaged convention; ES~Cet is marked by a star. Filled pale symbols show literature comparison systems compiled by \citet{Yu2026}. Curves show the TianQin, LISA, Taiji, and DECIGO characteristic noise; the LISA and Taiji curves include the Galactic confusion foreground. The highest-lying open circle is the newly identified 37-pc candidate J213957 (Table~\ref{tab:high_priority}). Grey bars give the $1\sigma$ distance uncertainty propagated to $h_c$ (vertical; $h_c\propto1/d$, taken from the Gaia parallax precision) and the conservative period uncertainty propagated to $\fGW$ (horizontal; smaller than the symbols except for the three multiband-poor sources of Section~\ref{sec:period_def}). They do not include the dominant uncertainty, the assumed chirp mass: since $h_c\propto\Mc^{5/3}$, a factor of 1.5 in $\Mc$ moves every open circle by a factor of 2. Candidate positions are fiducial estimates under the assumptions stated in Section~\ref{sec:gw} \citep{Robson2019,Yu2026}.
    \label{fig:gw_strain}}
\end{figure*}

Table~\ref{tab:gw} lists the fiducial SNRs for the DESI-characterized white-dwarf-cooling-sequence candidates plus ES~Cet \added{(eight of the nine; J214140 is omitted because its period is unreliable)}, adopting $\Mc=0.3\,\Msun$ and the averaged strain convention of Equation~(\ref{eq:strain}). Because SNR scales as $\Mc^{5/3}$, the tabulated values shift with the true chirp mass. Figure~\ref{fig:gw_strain} places all 147 candidates on the detector curves and adds the literature systems compiled by \citet{Yu2026} for comparison.

Within this spectroscopically characterized subset, ZTF~J071816.38+373138.66 (Section~\ref{sec:J071816}) has the largest fiducial SNR: assuming $\Porb=22.55$~min and $\fGW=1.479$~mHz at $d=85$~pc and $\Mc=0.3\,\Msun$ gives $h=2.18\times10^{-21}$, $\rho_{\rm TQ}=79.3$, and $\rho_{\rm LISA}=220.5$. J162009 and J000637 give $\rho_{\rm LISA}=29.3$ and 25.4; J152934, J100316, and J000208 also remain above $\rho_{\rm LISA}=5$.

Extending the same fiducial computation to the full catalog (Figure~\ref{fig:gw_strain}; per-source values are released as a supplementary table), 17 of the newly identified WD-locus candidates lie above $\rho_{\rm LISA}=5$, including the six DESI-characterized systems above. \added{None of the newly identified WD-locus candidates has a measured chirp mass, and $\rho\propto\Mc^{5/3}$: the count is 9 for $\Mc=0.15\,\Msun$, 17 for the adopted $0.3\,\Msun$, and 21 for $0.6\,\Msun$ (TianQin: 4, 11, and 20). Each count also assumes the modulation is orbital, which Section~\ref{sec:harmonic_test} shows is not established for the compact WD-locus majority. These are prioritization numbers under stated assumptions, not detection forecasts. Distances here are inverse parallaxes, adequate for the bulk of the 17 (median $\varpi/\sigma_\varpi=107$) but not for the one member at $\varpi/\sigma_\varpi=2.3$, whose distance, and hence signal-to-noise ratio, should be treated as unconstrained.} The formally strongest is ZTF~J213957.39$-$124550.08, the first entry of Table~\ref{tab:high_priority}: its $\Porb=21.26$~min and $d=36.7$~pc ($\varpi/\sigma_\varpi=461$) imply $\rho_{\rm LISA}\approx630$ under the fiducial assumptions, and it corresponds to the highest-lying open circle in Figure~\ref{fig:gw_strain}.

That figure is also the clearest illustration of how conditional these rankings are. At face value J213957 would be an order of magnitude louder than any confirmed verification binary and would merge within 5~Myr while sitting at 37~pc. It is in fact a well-studied member of the 40-pc white-dwarf sample \citep{Tremblay2020}, a DA of $\Teff=7914\pm130$~K and $\logg=8.39$ \citep{GentileFusillo2021}---an ordinary, slightly massive, single cooling white dwarf whose Gaia luminosity leaves no room for a comparably bright companion. Its modulation is significantly chromatic ($A_g/A_r=1.41\pm0.11$), like that of J162009, and its maximum possible ellipsoidal amplitude is $6\times10^{-5}$~mag against an observed $0.11$~mag (Section~\ref{sec:harmonic_test}). We therefore report it as the formally strongest conditional candidate and simultaneously as the one most in need of refutation; archival TESS photometry (TIC~242073510) can test the coherence and harmonic content of the 638-s signal directly.

The assumed frequencies of the DESI-matched candidates mostly lie below TianQin's peak sensitivity band ($f \approx 10$--$100$~mHz; \citealt{Luo2016}), so TianQin detectability depends strongly on distance, chirp mass, and mission duration. The TianQin+LISA network can nevertheless improve sky localization and parameter estimation for confirmed systems \citep{Yu2026}.  DECIGO, with its different design sensitivity, would provide much higher fiducial SNRs for the nearest systems under the same binary assumptions.

The 69 hot-subdwarf-region candidates are also relevant to the mHz population. Short-period sdB+WD binaries can be strong LISA/TianQin sources \citep{Kupfer2018,Kupfer2020}. Our candidates span $\Porb=23.05$--$106.46$~min (median 93.63~min), with five below 40~min. A genuine sdB+WD among the shortest-period systems would be potentially detectable by LISA, whereas an sdB+dM reflection binary would have a substantially smaller chirp mass.

GW-driven orbital decay is also observable electromagnetically through eclipse timing. For a circular binary the decay rate scales as $\dot{P}\propto-\Mc^{5/3}P^{-5/3}$ \citep{Peters1963}, and the accumulated timing shift is $\Delta t_{\rm O-C}\approx\frac12|\dot{P}/P|\,t^2$. For the deep eclipsing WD-locus source J160335, the catalog $\Porb=78.39$~min and the fiducial $\Mc=0.3\,\Msun$ give $\dot{P}\approx-0.0118$~ms~yr$^{-1}$ and a 5-s shift after approximately 11.2~yr (8.0~yr for $\Mc=0.45\,\Msun$). These are conditional GR-only estimates: the component masses and orbital interpretation remain unconfirmed, and the H$\alpha$ emission indicates that irradiation or accretion-related evolution may also affect the timing.

\subsection{Spatial and Period Distributions}
\label{sec:spatial_period}

Measured against the Galactic plane drawn in Figure~\ref{fig:sky}, the catalog is mildly concentrated toward the disk: the median $|b|$ is $20\fdg5$, against $31\fdg9$ for an isotropic distribution over the same $\delta>-28^\circ$ footprint, and half the sample lies at $|b|<20^\circ$ against 32\% isotropically. This is not evidence for a thin-disk population. The ZTF Galactic Plane Survey samples low latitudes more densely than the main survey, while Gaia crowding and the astrometric-quality criteria act in the opposite direction there, so the observed latitude distribution mixes both selection effects with any intrinsic one.

Consistent with their higher luminosities, the hot-subdwarf-region candidates are sampled at systematically larger distances and Galactic heights than the WD-locus candidates: restricting to $\varpi/\sigma_\varpi\geq5$ and computing $|z|=d\,|\sin b|$ with $d=1000/\varpi$~pc gives median $|z|\approx330$~pc (11 sources), versus 74~pc for the WD locus (38 sources) and 114~pc for the intermediate region (7 sources; Figure~\ref{fig:period_distribution}(b)). Because the catalog is neither volume limited nor selected uniformly in distance, this is a geometric consistency check, not a scale-height measurement.

The adopted $\Porb$ distribution differs by CMD class (Figure~\ref{fig:period_distribution}(a)). Of the 52 WD-locus sources, 15 have $\Porb<40$~min and the sample median is 70.37~min. The 69 hot-subdwarf-region sources have a median of 93.63~min, with 35/69 in the 90--100~min bin, 14 above 100~min, and only five below 40~min. The concentration at and just above the 100-min boundary is shaped by the $<100$~min search window on $\Pref$ and the source-by-source selection of the orbital harmonic; it is therefore not interpreted as an intrinsic period minimum. Hot-subdwarf light curves in this range can arise from reflection, eclipses, ellipsoidal modulation, pulsation, or rotation \citep{Wang2021}, so phase-resolved spectroscopy and higher-cadence multiband photometry are required to determine the physical origin. The 15 WD-locus candidates below 40~min remain the most direct targets for compact-binary and GW follow-up.

\subsection{Outlook for LSST and CSST}
\label{sec:outlook}

The present catalog is limited by Gaia variability selection, ZTF sampling, and the need for per-source period refinement and refolding of every ranked light curve. Rubin/LSST \citep{Ivezic2019} will extend this approach to fainter sources and a complementary southern footprint, while CSST \citep{Gong2019} will provide deep, high-resolution multiband imaging that can improve counterpart identification and composite-SED constraints. The recovery of 10--100~min periods will still depend on the delivered cadence, so we do not extrapolate a discovery yield from survey depth alone. The CNN ranking step is intended as a scalable pre-filter, whereas physical confirmation will continue to require time-resolved spectroscopy and high-cadence photometry.

\section{Summary}
\label{sec:summary}

We present a Gaia DR3 and ZTF DR23 catalog of 147 short-period blue compact-binary candidates. The principal results are:

\begin{enumerate}
\item Five Gaia selection criteria retain 187, 169, 166, 159, and 122 members of the 209-source reference set. All targets promoted by the ZTF period searches and CNN morphology ranking were refolded at high-precision refined periods; 147 candidates with coherent short-period modulation were retained. \added{A stage-by-stage budget (Table~\ref{tab:completeness}) attributes the end-to-end recovery of 28 of the 209 reference systems to the Gaia criteria (87 systems), the ZTF footprint and sampling (about 20), and the combined detection, ranking, and vetting stage (the remainder); the 3--100~min search window itself costs nothing, since every reference period falls inside it. The trials factor of the period search is expected to contribute of order one spurious entry to the 147 (Section~\ref{sec:sample}).}
\item Thirty-six catalog members are recovered compact binaries and 111 are newly identified compact-binary candidates; the composition by CMD region, morphology, novelty, and DESI coverage is summarized in Table~\ref{tab:composition}.
\item Twenty-six catalog members have DESI DR1 spectra. Gaia luminosities and spectral morphology separate 9 white-dwarf-cooling-sequence candidates, 14 luminous-blue systems, and 3 known or peculiar objects. Single-epoch DESI spectra do not establish double-white-dwarf orbits.
\item Under the fiducial compact-binary assumptions adopted here, 17 newly identified white-dwarf-locus candidates would exceed the adopted LISA threshold, in addition to ES~Cet; six belong to the DESI-characterized subset, and the formally strongest is J213957 ($\rho_{\rm LISA}\approx630$ at 37~pc). \added{No candidate has a measured chirp mass; with $\rho\propto\Mc^{5/3}$, the count is 9 for $\Mc=0.15\,\Msun$ and 21 for $\Mc=0.6\,\Msun$.} These rankings are conditional on the modulation being orbital, which Section~\ref{sec:harmonic_test} shows is not established: for the compact white-dwarf-locus majority the observed amplitudes exceed the maximum possible ellipsoidal signal by three to five orders of magnitude, and rotating magnetic or double-faced white dwarfs reproduce the light curves. Ultraviolet and mid-infrared color diagnostics for the 74-source SED subsample identify hot components---most clearly the optically featureless white dwarf ZTF~J071816---and flag two short-period candidates with significant $W1-W2$ excesses (Appendix~\ref{app:uvir}).
\item For $\varpi/\sigma_\varpi\geq5$, the WD-locus and hot-subdwarf-region subsets have median $|z|\approx74$ and $330$~pc, respectively, although the selection does not permit a scale-height measurement. Fifteen WD-locus candidates have adopted $\Porb<40$~min; ten of these are newly identified, and seven also have $\MG\geq11$ with $\varpi/\sigma_\varpi\geq5$ (Table~\ref{tab:high_priority}). Five of the seven have a published spectral type but none has a time-resolved radial-velocity series. These and the deep eclipsers define the highest-priority follow-up sample.
\end{enumerate}

Time-resolved radial-velocity spectroscopy is required to confirm the binary interpretation and measure component masses. The complete 147-source catalog---with periods, Gaia classifications, and spectroscopic flags---is released in machine-readable form with this paper, together with supplementary tables of the SED-subsample multiwavelength photometry and the per-source fiducial gravitational-wave quantities, as a follow-up resource for electromagnetic and gravitational-wave studies.

\section*{Data and Code Availability}

\added{Three machine-readable tables accompany this paper, each with a printed example in the article.} The online material contains the final 147-source catalog with coordinates, adopted $\Porb$ and conservative period errors, EA/EW-like morphology, Gaia photometry and CMD region, prior-catalog status, and DESI coverage; \added{its columns are documented with the released table, and Table~\ref{tab:high_priority} prints an excerpt. It is the data behind Figures~\ref{fig:sky}, \ref{fig:hr}, and \ref{fig:period_distribution}}. Two supplementary machine-readable tables provide the per-source fiducial gravitational-wave quantities of all 147 candidates (assumed $\Porb$, distance, $\fGW$, $h_c$, and fiducial TianQin/LISA signal-to-noise ratios; Section~\ref{sec:gw}\added{; excerpt in Table~\ref{tab:gw}; the data behind Figure~\ref{fig:gw_strain}}) and the ultraviolet and mid-infrared photometry and colors of the 74-source SED subsample (Appendix~\ref{app:uvir}\added{; excerpt in Table~\ref{tab:mrt3}; the data behind Figure~\ref{fig:uvir}}). \added{The full multiband SED compilation from which the ultraviolet and mid-infrared colors are formed, the harmonic and ellipsoidal audit of all 147 sources, the rotation diagnostics of Table~\ref{tab:rotation}, the stage-by-stage flags behind Table~\ref{tab:completeness}, and the literature reference set used to calibrate the Gaia thresholds are released as supporting comma-separated files, with accompanying documentation defining every column and unit, in the Zenodo deposit \dataset[doi:10.5281/zenodo.21980866]{\doi{10.5281/zenodo.21980866}}. The morphology-ranking step uses the MobileNetV2 reference implementation cited in Section~\ref{sec:ml_samples} without architectural modification; the training set, class definitions, and held-out performance are specified there and in Figure~\ref{fig:cnn}.} The three rejected BLS-only objects are not included in any catalog or supplementary product. All manuscript tables and figures are generated from the same catalog and release version. The 464-object ranked-target list is not released as a scientific catalog. Gaia DR3, ZTF DR23, DESI DR1, SIMBAD, VizieR, MAST, and AllWISE are public resources.

\begin{acknowledgments}
J.L.\ and C.L.\ are supported by the National Natural Science Foundation of China (NSFC grant No.~12233013). Y.H.\ and Y.L.\ are supported by the National Key Research and Development Program of China (No.~2023YFC2206701). This work is funded by CMS-CSST-2025-A13, ``Large Sample Search and Study of Late-Stage Stellar Evolution Objects'', from the Chinese Space Station Telescope (CSST). We acknowledge the Gaia mission (\url{https://www.cosmos.esa.int/gaia}), processed by the Gaia Data Processing and Analysis Consortium (DPAC). Based on observations obtained with the Samuel Oschin Telescope 48-inch and the 60-inch Telescope at the Palomar Observatory as part of the Zwicky Transient Facility project. ZTF is supported by the National Science Foundation under Grants No.\ AST-1440341 and AST-2034437 and a collaboration including current partners Caltech, IPAC, the Oskar Klein Center at Stockholm University, the University of Maryland, University of California, Berkeley, the University of Wisconsin at Milwaukee, University of Warwick, Ruhr University, Cornell University, Northwestern University and Drexel University. Operations are conducted by COO, IPAC, and UW. This research used data obtained with the Dark Energy Spectroscopic Instrument (DESI). DESI construction and operations is managed by the Lawrence Berkeley National Laboratory. This material is based upon work supported by the U.S. Department of Energy, Office of Science, Office of High-Energy Physics, under Contract No.\ DE-AC02-05CH11231, and by the National Energy Research Scientific Computing Center, a DOE Office of Science User Facility under the same contract. Additional support for DESI was provided by the U.S. National Science Foundation (NSF), Division of Astronomical Sciences under Contract No.\ AST-0950945 to the NSF's National Optical-Infrared Astronomy Research Laboratory; the Science and Technology Facilities Council of the United Kingdom; the Gordon and Betty Moore Foundation; the Heising-Simons Foundation; the French Alternative Energies and Atomic Energy Commission (CEA); the National Council of Humanities, Science and Technology of Mexico (CONAHCYT); the Ministry of Science and Innovation of Spain (MICINN), and by the DESI Member Institutions: \url{https://www.desi.lbl.gov/collaborating-institutions}. The DESI collaboration is honored to be permitted to conduct scientific research on I'oligam Du'ag (Kitt Peak), a mountain with particular significance to the Tohono O'odham Nation. This research has made use of SIMBAD and VizieR, operated at CDS, Strasbourg, France; MAST; and the AllWISE catalog.
\end{acknowledgments}

\facilities{Gaia, PO:1.2m (ZTF), Mayall (DESI), GALEX, WISE, Sloan, 2MASS}

\software{Astropy \citep{Astropy2022}, matplotlib \citep{matplotlib2007}, numpy \citep{numpy2020}, pandas \citep{pandas2020}, P4J \citep{P4J2018}, \texttt{cuvarbase} (\url{https://github.com/johnh2o2/cuvarbase}), TensorFlow \citep{tensorflow2016}, MobileNetV2 \citep{Sandler2018}}


\clearpage
\appendix

\section{Gaia Selection Criteria: Equations and Diagnostics}
\label{app:gaia_criteria}

This appendix records the equation-level definitions and diagnostic figures for Criteria~2--5 summarized in Section~\ref{sec:gaia_filter} and Table~\ref{tab:gaia_criteria}. Criterion~1 remains in the main text (Section~\ref{sec:cmd}) because it defines the blue compact-object locus used throughout the paper.

\subsection{Criterion 2: Measurement Quality Control}
\label{app:quality}

To ensure reliable astrometric and photometric measurements, we apply eight quality thresholds optimized via the ratio method:
\beq
\texttt{parallax\_over\_error} > 0.5 \label{eq:q1}
\eeq
\beq
\texttt{phot\_g\_mean\_mag} > 11.979 \label{eq:q2}
\eeq
\beq
\texttt{astrometric\_sigma5d\_max} < 2.14 \label{eq:q3}
\eeq
\beq
\texttt{visibility\_periods\_used} > 9.85 \label{eq:q4}
\eeq
\beq
\texttt{astrometric\_excess\_noise} < 1 \label{eq:q5}
\eeq
\beq
\texttt{phot\_rp\_mean\_flux\_over\_error} > 5.7 \label{eq:q6}
\eeq
\beq
\texttt{phot\_bp\_mean\_flux\_over\_error} > 4.45 \label{eq:q7}
\eeq
\beq
\texttt{phot\_g\_mean\_flux\_over\_error} > 9 \label{eq:q8}
\eeq
The bright limit (Equation~\ref{eq:q2}) excludes saturated sources, while the flux S/N thresholds (Equations~\ref{eq:q6}--\ref{eq:q8}) guard against spurious variability induced by photon noise. After this step, 98,854,486 sources remain, retaining 169 of the 209 reference systems (80.9\% cumulative; 90.4\% of the 187 entering this criterion).

\subsection{Criterion 3: Proper-Motion/Parallax Index Cut}
\label{app:rpm}

We use the standard reduced proper motion \citep{Luyten1922},
\beq
H_G = G + 5\log_{10}(\mu_{\rm mas/yr}) - 10,
\eeq
where $\mu=(\mu_\alpha^2+\mu_\delta^2)^{1/2}$ is the total proper motion in mas~yr$^{-1}$. This kinematic luminosity proxy helps separate intrinsically faint, high-proper-motion populations from main-sequence stars at the same color. Calibrating the threshold by the same retention-rate ratio used for the other criteria, we adopt
\beq
H_G > 5.3\,(G_{\rm BP}-G_{\rm RP}) + 5.9
\label{eq:rpm}
\eeq
in the Gaia DR3 photometric system. This criterion reduces the sample to 73,774,968 sources while retaining 166 reference systems (79.4\% cumulative; 98.2\% of the 169 that survived Criterion~2).

\subsection{Criterion 4: Astrometric UWE and Photometric Consistency}
\label{app:uwe}

\begin{figure*}[htbp]
\centering
\includegraphics[width=0.95\textwidth]{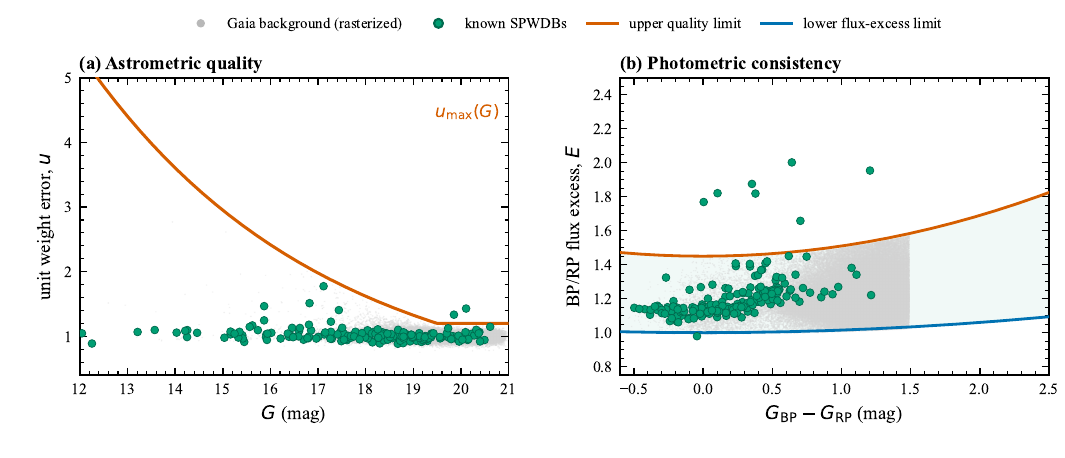}
\caption{Gaia quality filtering for Criterion~4. \textit{(a)} Astrometric unit weight error, $u=\sqrt{\chi^2_{\rm al}/(N_{\rm al}-5)}$, versus $G$ magnitude; the orange curve is the adopted upper limit. \textit{(b)} BP/RP flux-excess factor versus $G_{\rm BP}-G_{\rm RP}$; the blue and orange curves are the adopted lower and upper limits, respectively. Grey points show the Gaia field population, green symbols the 166 reference systems entering Criterion~4, and the shaded regions the accepted parameter space.}
\label{fig:gaia_quality}
\end{figure*}

We apply an astrometric unit weight error (UWE) upper limit to exclude sources with poor astrometric solutions, following \citet{Lindegren2018}:
\beq
u < 1.2 \times \max\!\bigl(1,\;e^{-0.2\,(\texttt{phot\_g\_mean\_mag}-19.5)}\bigr)
\label{eq:uwe}
\eeq
where $u = \sqrt{\chi^2_{\rm al}/(N_{\rm al}-5)}$, with $\chi^2_{\rm al}$ the along-scan astrometric chi-squared and $N_{\rm al}$ the number of good along-scan observations. Following \citet{Pelisoli2018}, who showed this criterion effectively removes $\gtrsim$97\% of sources contaminated by nearby companions while retaining known ELM white dwarfs, we require photometric consistency:
\beq
1.0+0.015\,(G_{\rm BP}-G_{\rm RP})^2 < E < 1.45+0.06\,(G_{\rm BP}-G_{\rm RP})^2
\label{eq:phot_consist}
\eeq
where $E = (I_{\rm BP}+I_{\rm RP})/I_G$ is the BP/RP flux excess factor. Sources with anomalous $E$ at a given color indicate blending with a nearby source, background contamination, or an unresolved binary with an extreme flux ratio. Figure~\ref{fig:gaia_quality} shows the two quality cuts for the 166 reference systems entering Criterion~4, together with the Gaia field population; 159 reference systems satisfy both requirements (76.1\% cumulative).

\subsection{Criterion 5: Variability Index}
\label{app:varindex}

To identify sources with significant photometric variability consistent with short-period binary modulation, we compute the Gaia variability metric following \citet{Guidry2021} and \citet{ElBadry2021b}:
\beq
\texttt{VARINDEX} = \frac{\sigma_G}{\langle G\rangle}\sqrt{n_{\rm obs}} - \bigl(A\,e^{\alpha\langle G\rangle} + B\,e^{\langle G\rangle - 17.0} + C\bigr)
\label{eq:varindex}
\eeq
where $\sigma_G = \langle G\rangle_{\rm flux}/({\tt phot\_g\_mean\_flux\_over\_error}\,)$ is the flux-scatter proxy recovered from the Gaia $G$ photometry and $n_{\rm obs}={\tt phot\_g\_n\_obs}$. The coefficients $A = 8.31\times10^{-9}$, $\alpha = 0.794$, $B = 0.0005$, and $C = 0.00962$ were refitted to the Gaia DR3 photon-noise floor over the color--magnitude region selected here, following the functional form of \citet{Guidry2021}. Sources with $\texttt{VARINDEX} > 0$ exhibit variability in excess of pure photon noise.

\begin{figure*}[htbp]
\centering
\includegraphics[width=0.95\textwidth]{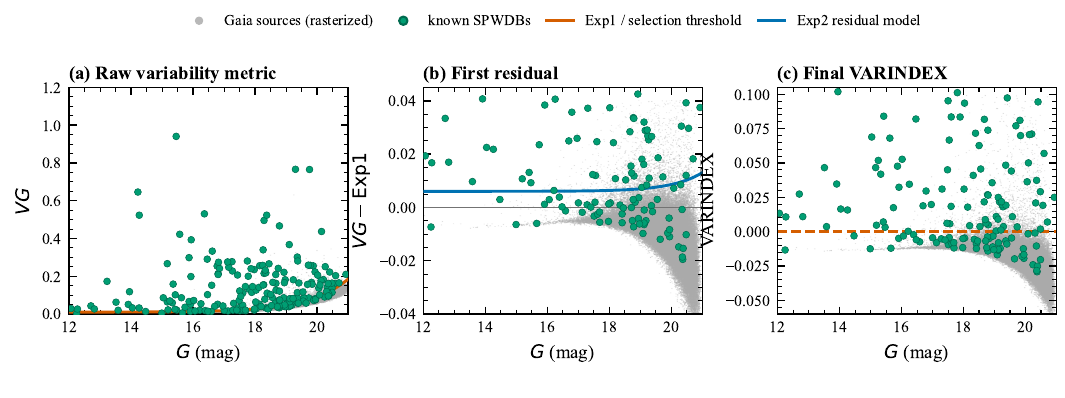}
\caption{Construction of the \texttt{VARINDEX} photometric-variability indicator (Criterion~5). Gray points show the Gaia source distribution and green circles the reference systems. \textit{(a)} The raw metric $VG=\sigma_G\sqrt{n_{\rm obs}}/\langle G\rangle$ and its first exponential photon-noise model. \textit{(b)} The residual after subtracting that model and the second exponential component. \textit{(c)} The final $\texttt{VARINDEX}=VG-\mathrm{Exp1}-\mathrm{Exp2}$; the dashed line marks the adopted threshold, $\texttt{VARINDEX}>0$.}
\label{fig:varindex}
\end{figure*}

We adopt the inclusive threshold $\texttt{VARINDEX}>0$. This reduces the approximately 66.4 million sources entering Criterion~5 to approximately 280,000 ZTF targets while retaining 122 of the 209 reference systems (58.4\% cumulative; 76.7\% of the 159 entering this criterion). The 37 lost reference objects are predominantly faint systems for which Gaia flux scatter is dominated by photon noise. Figure~\ref{fig:varindex} shows the construction of the variability indicator and the adopted threshold.

\section{Spectroscopic and Photometric Panels for the DESI-Matched Catalog Sources}
\label{app:desi}
\label{app:panels}

\begin{figure*}
\centering
\includegraphics[width=0.98\textwidth,height=0.86\textheight,keepaspectratio]{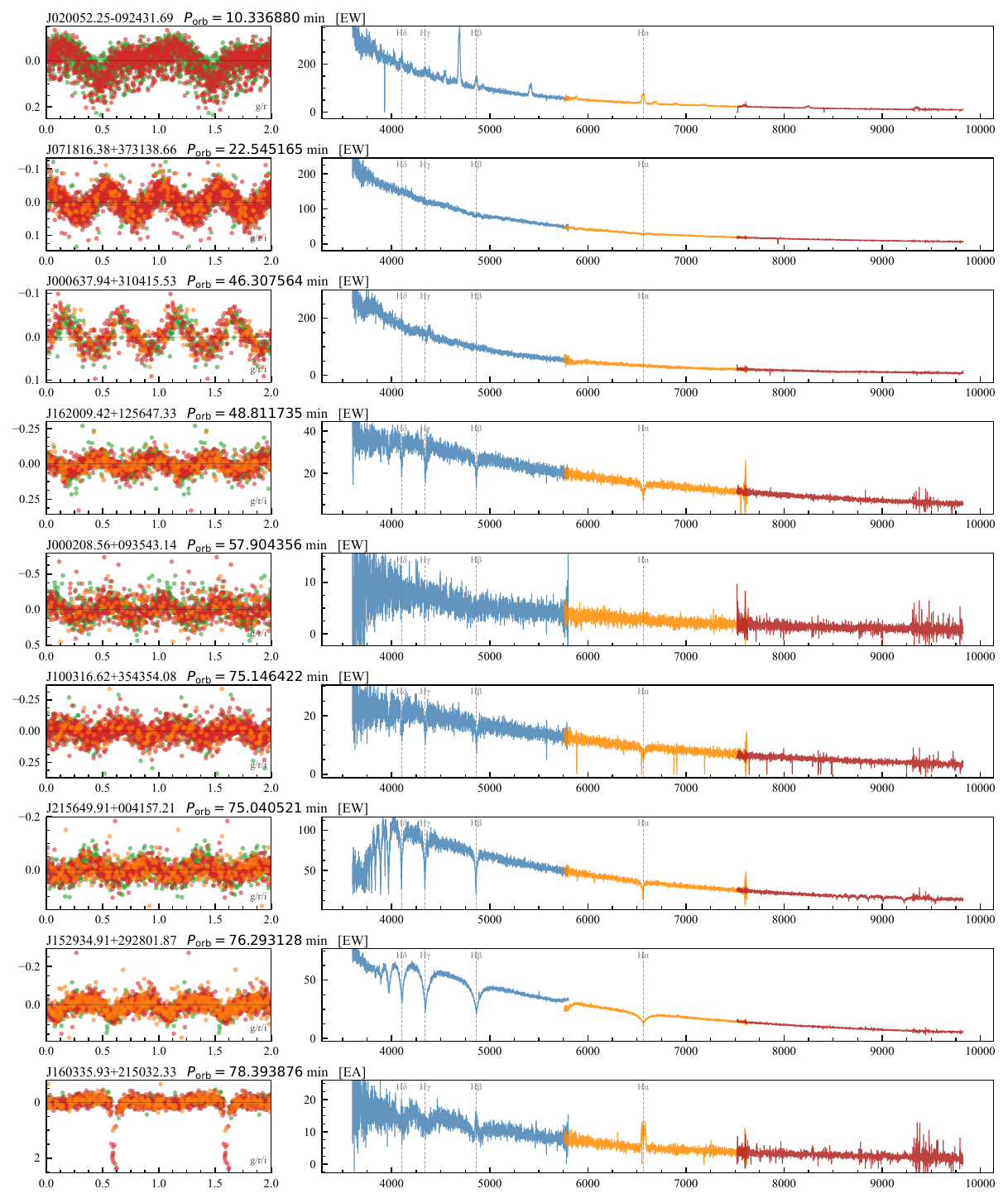}
\caption{Combined ZTF light-curve and DESI-spectrum panels for catalog sources \#1--9 with DESI coverage. Phase calculations and annotations use the full-precision catalog period. The top row is ES~Cet, whose fold is single-waved ($\Porb=\Pref=10.336880$~min); the deep eclipser J160335 is the bottom row.}
\label{fig:panelB1}
\end{figure*}

\begin{figure*}
\centering
\includegraphics[width=0.98\textwidth,height=0.86\textheight,keepaspectratio]{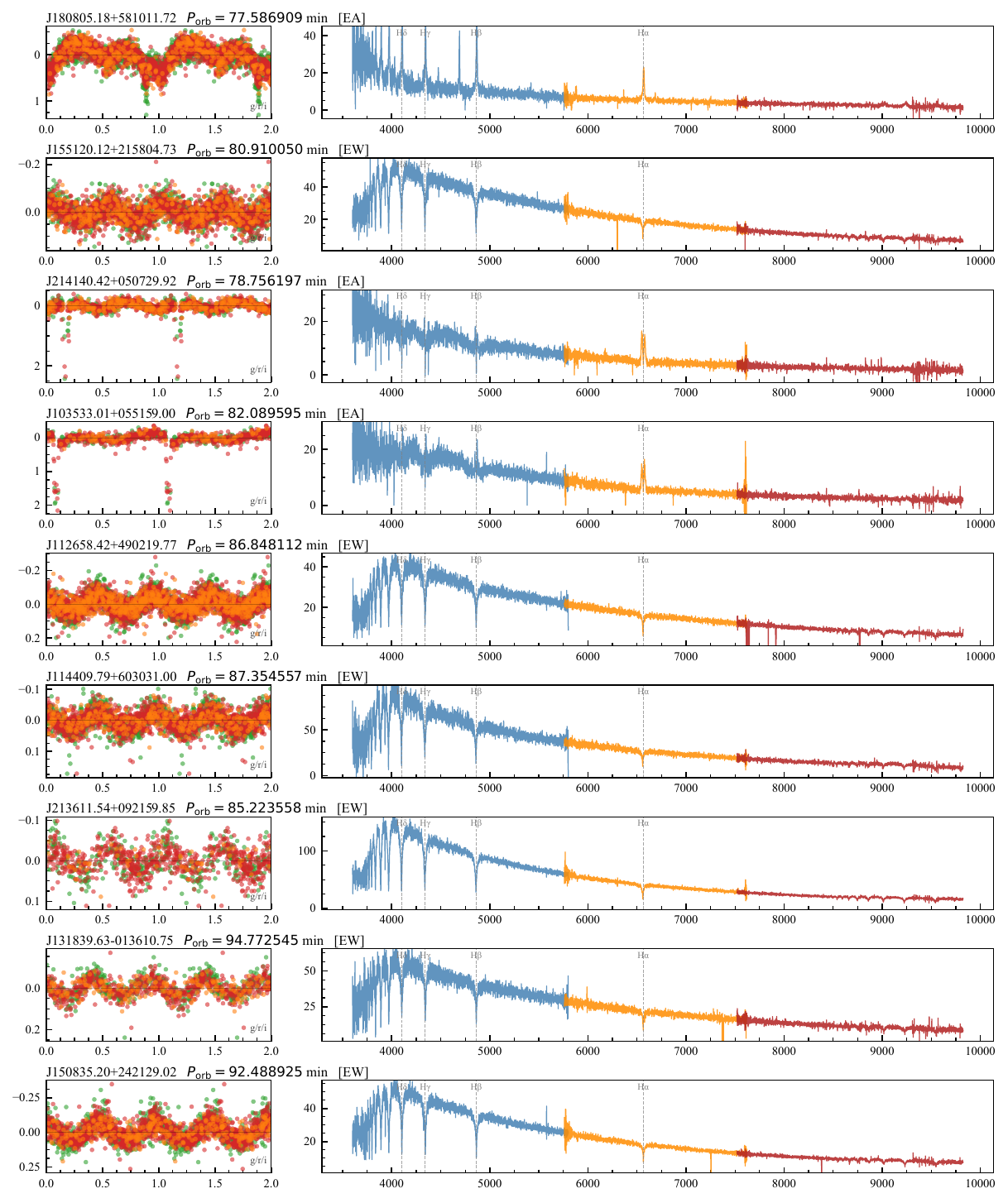}
\caption{Same as Figure~\ref{fig:panelB1}, for DESI-matched catalog sources \#10--18. J103533 and J214140 are the clearest deep eclipsers, and J180805 (1RXS~J180804.3+581001) shows emission-like Balmer features.}
\label{fig:panelB2}
\end{figure*}

\begin{figure*}
\centering
\includegraphics[width=0.98\textwidth,height=0.86\textheight,keepaspectratio]{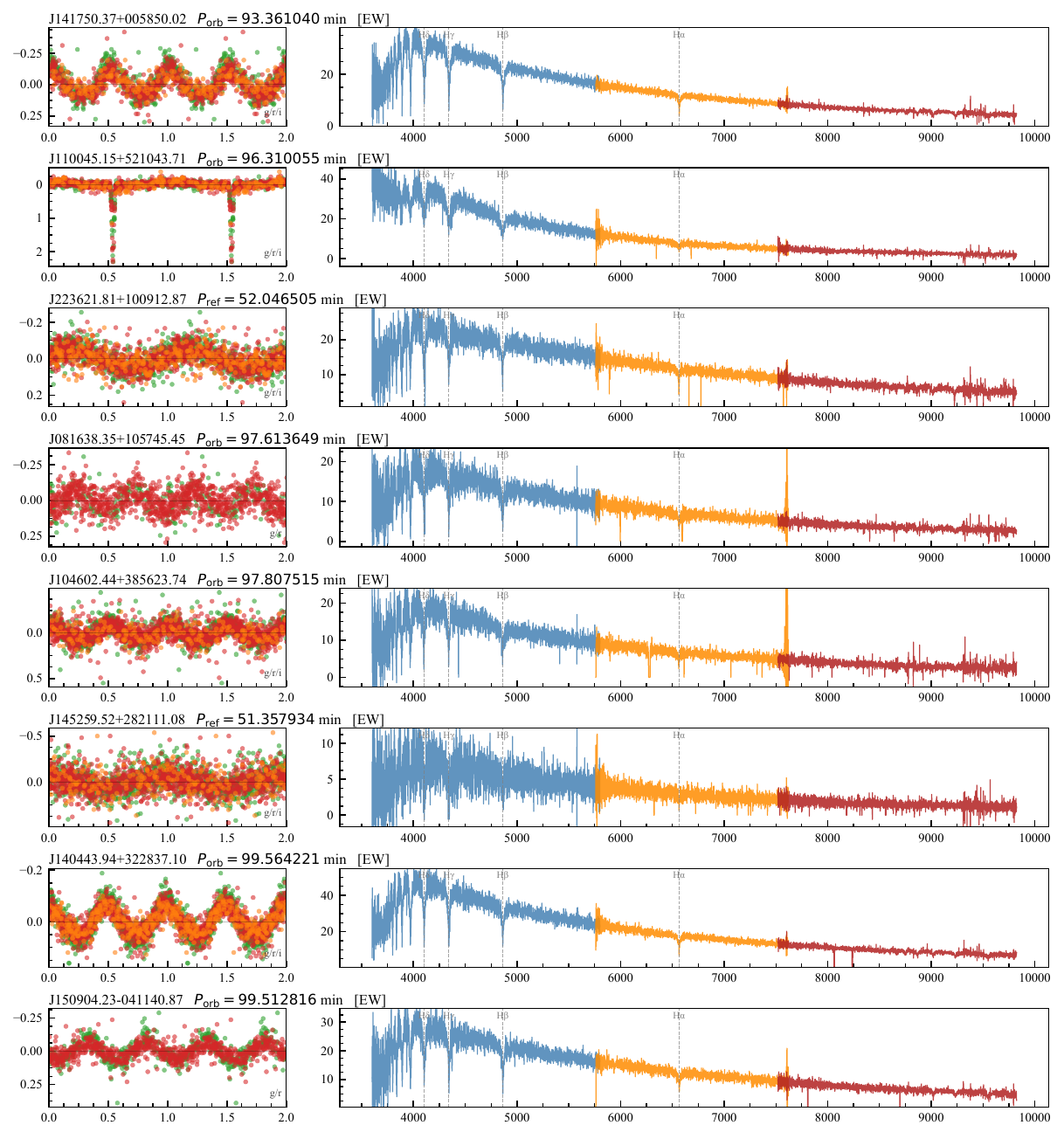}
\caption{Same as Figure~\ref{fig:panelB1}, for DESI-matched catalog sources \#19--26 (S/N$_R=9$--$31$). Most show Balmer absorption but are luminous-blue hot-subdwarf/BHB/A-type candidates rather than white dwarfs (Section~\ref{sec:new_desi_sources}); the exception is the \citet{Ren2023} compact binary ZTF~J110045.15+521043.71. J223621 and J145259 are folded at $\Pref$ (Section~\ref{sec:harmonic_test}); their adopted $\Porb$ in Table~\ref{tab:desi} is twice the value shown.}
\label{fig:panelB3}
\end{figure*}

Figures~\ref{fig:panelB1}--\ref{fig:panelB3} present combined light-curve and spectrum panels for the 26 DESI members of the 147-source catalog (Table~\ref{tab:desi}; adopted $\Porb=10.34$--$104.09$~min, with only ES~Cet having an independently established orbital period). All light curves are folded at the full-precision period at which the light curve repeats, and each panel title reports that value with its symbol: the adopted $\Porb$ for 24 sources, and $\Pref$ for J223621 and J145259, whose periodograms carry no significant power at the fundamental $1/\Porb$ (Section~\ref{sec:harmonic_test}), so a fold at $2\Pref$ would show the same cycle twice. Full precision matters here: at $3\times10^{5}$ cycles over the ZTF baseline, folding ES~Cet on a period rounded to two decimals drifts the phase by $\sim$150 cycles and erases the modulation entirely. Vertical dotted lines mark the Balmer series. The three rejected BLS-only objects are not included.

\begin{deluxetable*}{lcccccccc}
\tablewidth{0pt}
\tabletypesize{\scriptsize}
\tablecaption{Summary Properties of the 26 Compact-Catalog DESI Matches\label{tab:desi}}
\tablehead{
  \colhead{ZTF Name} & \colhead{Type} & \colhead{$\Porb$} & \colhead{\rev{$\Delta P_{\rm cons}$}} & \colhead{$G$} & \colhead{$d_{\rm geo}$} & \colhead{$M_G$} & \colhead{S/N$_R$} & \colhead{Class} \\
  \colhead{} & \colhead{} & \colhead{(min)} & \colhead{\rev{(min)}} & \colhead{(mag)} & \colhead{(pc)} & \colhead{(mag)} & \colhead{} & \colhead{}
}
\startdata
J020052.25$-$092431.69\tablenotemark{a} & EW & 10.3369 & 0.000016 & 16.80 & 1787 & 5.54 & 34.4 & special \\
J071816.38$+$373138.66 & EW & 22.5452 & 0.000037 & 16.95 & 85 & 12.32 & 49.1 & WD-CS \\
J000637.94$+$310415.53 & EW & 46.3076 & 0.000171 & 16.81 & 98 & 11.84 & 15.5 & WD-CS \\
J162009.42$+$125647.33 & EW & 48.8117 & 0.000174 & 18.08 & 75 & 13.69 & 18.7 & WD-CS \\
J000208.56$+$093543.14 & EW & 57.9044 & 0.000250 & 19.89 & 249 & 12.91 & 4.5 & WD-CS \\
J215649.91$+$004157.21 & EW & 75.0405 & 0.000416 & 17.10 & 2479 & 5.13 & 28.9 & off-sequence \\
J100316.62$+$354354.08 & EW & 75.1464 & 0.000408 & 18.54 & 107 & 13.39 & 12.7 & WD-CS \\
J152934.91$+$292801.87 & EW & 76.2931 & 0.000427 & 17.49 & 87 & 12.79 & 55.8 & WD-CS \\
J180805.18$+$581011.72\tablenotemark{b} & EA & 77.5869 & 0.000866 & 19.17 & 2878 & 6.88 & 7.5 & special \\
J160335.93$+$215032.33 & EA & 78.3939 & 0.000901 & 19.12 & 308 & 11.68 & 8.5 & WD-CS \\
J214140.42$+$050729.92 & EA & 78.7562 & 0.976488 & 19.00 & 274 & 11.81 & 7.2 & WD-CS \\
J155120.12$+$215804.73 & EW & 80.9101 & 0.000480 & 17.76 & 3531 & 5.02 & 22.5 & off-sequence \\
J103533.01$+$055159.00 & EA & 82.0896 & 0.000976 & 18.78 & 197 & 12.31 & 8.7 & WD-CS \\
J213611.54$+$092159.85 & EW & 85.2236 & 0.000534 & 16.88 & 4213 & 3.75 & 30.8 & off-sequence \\
J112658.42$+$490219.77 & EW & 86.8481 & 0.000544 & 17.98 & 4047 & 4.95 & 23.4 & off-sequence \\
J114409.79$+$603031.00 & EW & 87.3546 & 0.000550 & 17.25 & 4400 & 4.03 & 15.6 & off-sequence \\
J150835.20$+$242129.02 & EW & 92.4889 & 0.000628 & 17.96 & 3701 & 5.11 & 29.5 & off-sequence \\
J141750.37$+$005850.02 & EW & 93.3610 & 0.000650 & 18.29 & 2771 & 6.08 & 20.2 & off-sequence \\
J131839.63$-$013610.75 & EW & 94.7725 & 0.000679 & 17.63 & 3280 & 5.05 & 18.8 & off-sequence \\
J110045.15$+$521043.71\tablenotemark{c} & EW & 96.3101 & 0.000672 & 18.53 & 653 & 9.46 & 9.5 & special \\
J081638.35$+$105745.45 & EW & 97.6136 & 6.497928 & 18.95 & 1918 & 7.54 & 9.0 & off-sequence \\
J104602.44$+$385623.74 & EW & 97.8075 & 0.000690 & 18.96 & 2213 & 7.23 & 8.6 & off-sequence \\
J150904.23$-$041140.87 & EW & 99.5128 & 0.000737 & 18.40 & 6739 & 4.26 & 14.1 & off-sequence \\
J140443.94$+$322837.10 & EW & 99.5642 & 0.000738 & 17.71 & 3372 & 5.07 & 18.9 & off-sequence \\
J145259.52$+$282111.08 & EW & 102.7161 & 1.650350 & 19.70 & 1552 & 8.74 & 4.9 & off-sequence \\
J223621.81$+$100912.87 & EW & 104.0930 & 0.000798 & 18.41 & 3040 & 6.00 & 13.4 & off-sequence \\

\enddata
\tablecomments{$\Porb$ is the adopted photometric orbital-period estimate and \rev{$\Delta P_{\rm cons}$ its conservative uncertainty (Section~\ref{sec:period_def})}. Type is the EA/EW-like morphology. Distances are Bailer--Jones geometric distances; the fractional distance (and hence $\MG$) uncertainty is set by the parallax, $\sigma_{M_G}\approx2.17\,(\varpi/\sigma_\varpi)^{-1}$~mag, and the per-source $\varpi/\sigma_\varpi$ is included in the released catalog. Class denotes a white-dwarf cooling-sequence source (WD-CS), a luminous-blue source too bright for a white dwarf at its Gaia distance, or a known/peculiar system.}
\tablenotetext{a}{ES~Cet, a confirmed AM~CVn verification binary; we adopt our refined ZTF period, $\Porb=10.3369$~min, which agrees with the published 620.2-s orbit \citep{Espaillat2005}.}
\tablenotetext{b}{1RXS~J180804.3$+$581001; X-ray/CV-like emission spectrum.}
\tablenotetext{c}{ZTF~J1100$+$5210; recovered \citet{Ren2023} compact binary.}
\end{deluxetable*}

\section{High-Precision Phase-Folded Light Curves without DESI Spectra}
\label{app:lconly}

\begin{figure*}
\centering
\includegraphics[width=0.98\textwidth,height=0.88\textheight,keepaspectratio]{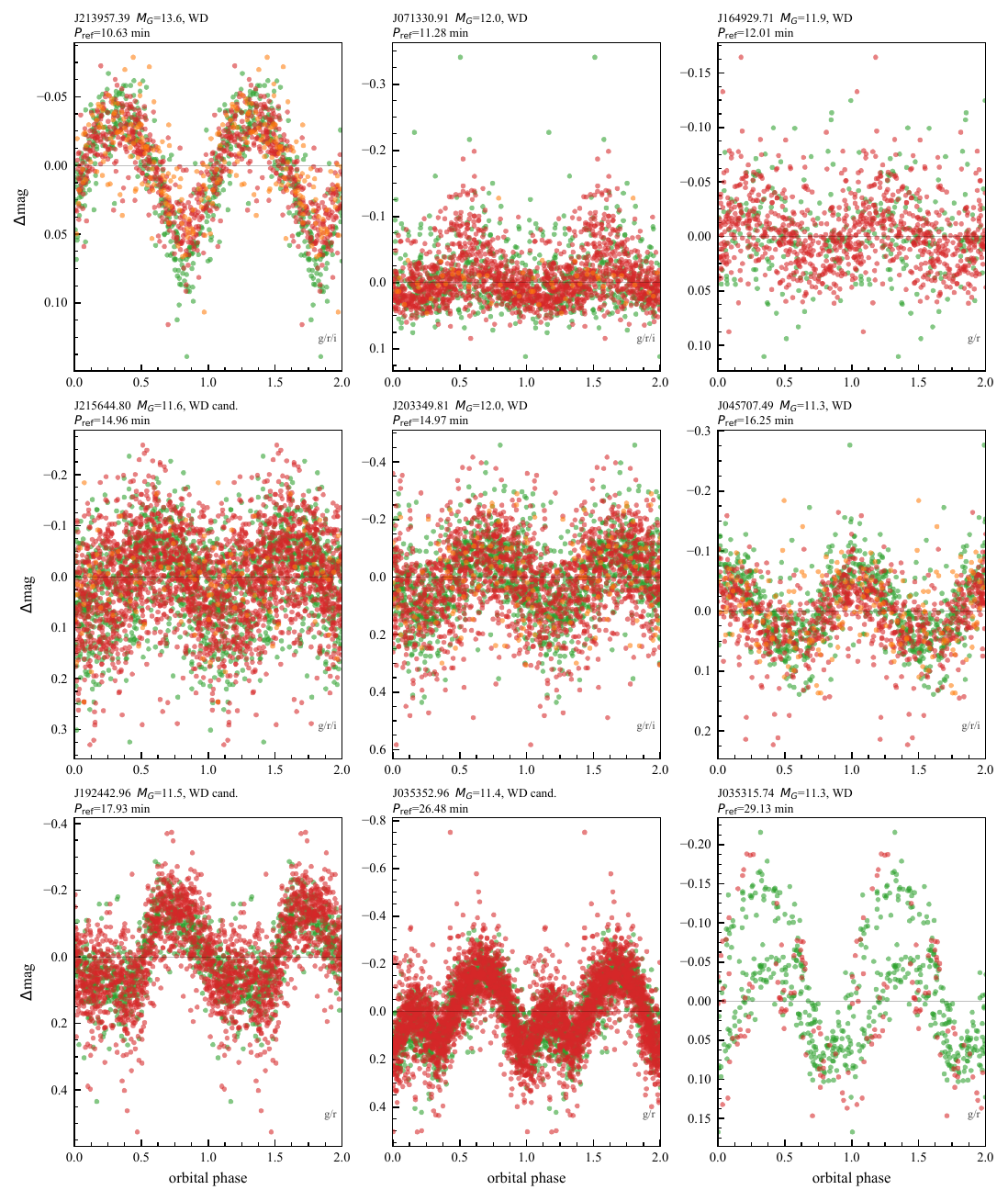}
\caption{High-precision phase-folded ZTF light curves for non-DESI candidates \#1--9 with $\MG\geq11$\added{, ordered by period. Small points are the individual ZTF epochs and filled symbols the phase-binned medians.} \added{Each panel is folded at the adopted catalog $\Pref$; the panels of the submitted version used an independent refinement that had locked onto the sidereal-day alias for nine of the eighteen sources.}}
\label{fig:lconly1}
\end{figure*}

\begin{figure*}
\centering
\includegraphics[width=0.98\textwidth,height=0.88\textheight,keepaspectratio]{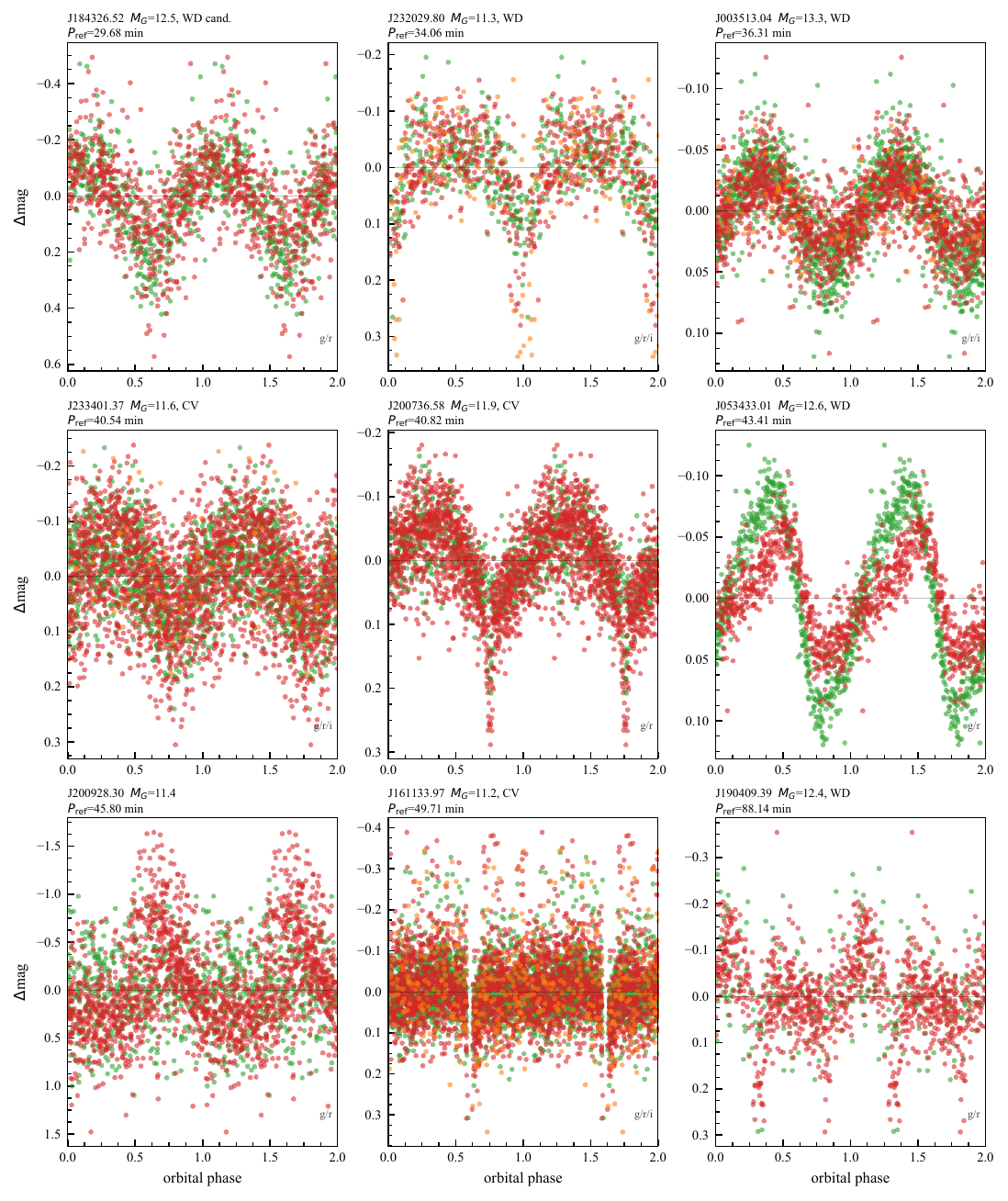}
\caption{Same as Figure~\ref{fig:lconly1}, for non-DESI candidates \#10--18.}
\label{fig:lconly2}
\end{figure*}

Figures~\ref{fig:lconly1}--\ref{fig:lconly2} present the high-precision phase-folded ZTF light curves for the 18 non-DESI members with $\MG\geq11$; Table~\ref{tab:lconly} lists their adopted periods, Gaia photometry, and SIMBAD types. Each panel uses its source-specific repeating period $\Pref$, \added{taken from the released catalog so that the gallery, the printed table and the machine-readable table agree}. $\Delta$ magnitudes are plotted relative to each band's median over two displayed cycles. 

\begin{deluxetable*}{lcccccl}
\tablewidth{0pt}
\tabletypesize{\scriptsize}
\tablecaption{Adopted periods for the non-DESI candidates with $\MG\geq11$\label{tab:lconly}}
\tablehead{
  \colhead{ZTF Name} & \colhead{$\Pref$} & \colhead{$\Delta P_{\rm cons}$} & \colhead{$G$} & \colhead{$\MG$} & \colhead{Type} & \colhead{SIMBAD} \\
  \colhead{} & \colhead{(min)} & \colhead{(min)} & \colhead{(mag)} & \colhead{(mag)} & \colhead{} & \colhead{}
}
\startdata
J213957.39$-$124550.08 & 10.628531 & 0.000017 & 16.41 & 13.58 & EW & WhiteDwarf \\
J071330.91$-$012623.33 & 11.280150 & 0.289353 & 16.84 & 11.96 & EW & WhiteDwarf \\
J164929.71$-$243310.22 & 12.012165 & 0.850872 & 17.15 & 11.88 & EW & WhiteDwarf \\
J215644.80$+$613633.37 & 14.958797 & 0.000032 & 18.99 & 11.60 & EW & WhiteDwarf Candidate \\
J203349.81$+$322901.10 & 14.967050 & 0.000032 & 20.01 & 11.96 & EW & WhiteDwarf \\
J045707.49$+$051322.03 & 16.249977 & 0.000040 & 18.31 & 11.27 & EW & WhiteDwarf \\
J192442.96$+$310403.62 & 17.934506 & 0.000046 & 19.37 & 11.46 & EW & WhiteDwarf Candidate \\
J161133.97$+$630831.66 & 49.708173 & 0.891134 & 18.29 & 11.24 & EA & CataclyV* \\
J035352.96$+$431525.16 & 26.480876 & 0.000105 & 19.16 & 11.36 & EW & WhiteDwarf Candidate \\
J035315.74$+$095633.40 & 29.133232 & 0.574590 & 17.07 & 11.30 & EW & WhiteDwarf \\
J184326.52$+$185021.24 & 29.677716 & 0.000127 & 19.93 & 12.53 & EW & WhiteDwarf Candidate \\
J232029.80$-$175734.64 & 34.055676 & 0.000173 & 18.04 & 11.33 & EW & WhiteDwarf \\
J003513.04$-$122511.26 & 36.314769 & 0.000197 & 16.89 & 13.28 & EW & WhiteDwarf \\
J233401.37$+$392138.64 & 40.542568 & 0.000242 & 16.03 & 11.63 & EW & CataclyV* \\
J200736.58$+$174214.34 & 40.815252 & 0.000241 & 15.18 & 11.90 & EW & CataclyV* \\
J053433.01$+$770755.05 & 43.407947 & 0.000272 & 16.51 & 12.60 & EW & WhiteDwarf \\
J190409.39$-$274049.68 & 88.138622 & 3.071397 & 18.62 & 12.42 & EA & WhiteDwarf \\
J200928.30$+$120204.39 & 45.798264 & 2.079541 & 19.83 & 11.45 & EW & -- \\

\enddata
\tablecomments{The 18 non-DESI catalog members with Gaia $\MG\geq11$. $\Pref$ is the high-precision repeating period used in Figures~\ref{fig:lconly1}--\ref{fig:lconly2}, and $\Delta P_{\rm cons}$ its conservative uncertainty (Section~\ref{sec:period_def}). Type is the EA/EW-like morphology, and SIMBAD records the pre-existing broad object type where available.}
\end{deluxetable*}

\section{Ultraviolet and Mid-Infrared Photometry of the SED Subsample}
\label{app:uvir}

\begin{figure*}[htbp]
\centering
\includegraphics[width=0.92\textwidth]{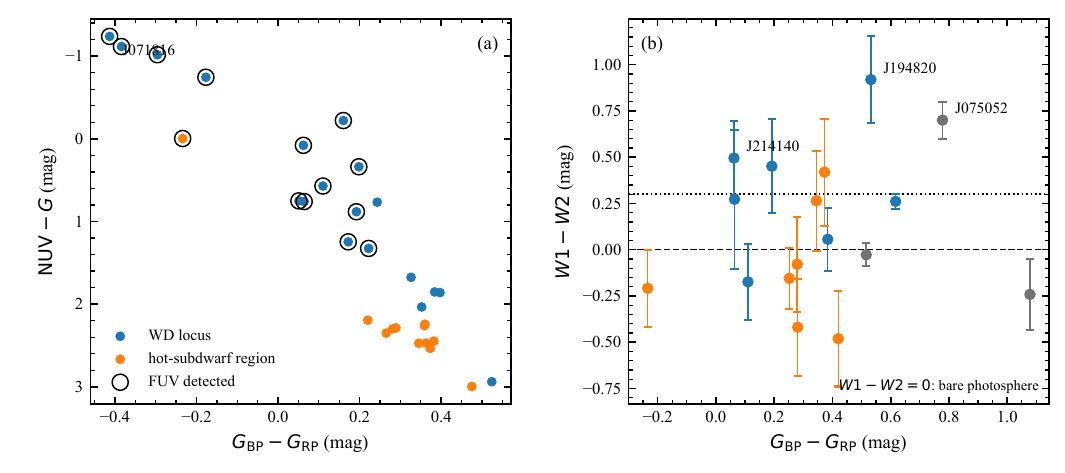}
\caption{Ultraviolet and mid-infrared color diagnostics for the SED subsample, colored by Gaia CMD region. \textit{(a)} ${\rm NUV}-G$ versus $G_{\rm BP}-G_{\rm RP}$ for the 31 GALEX NUV detections (S/N$\geq$3); black rings mark the 14 FUV detections, and the bluest source, J071816, is labeled. \textit{(b)} AllWISE $W1-W2$ versus $G_{\rm BP}-G_{\rm RP}$ for the 17 sources with reliable detections in both bands (\texttt{ph\_qual} A/B, clean contamination flags, single blend component). The dashed line marks the bare-photosphere expectation and the dotted line the $+0.3$~mag excess threshold; the three significant-excess sources are labeled.}
\label{fig:uvir}
\end{figure*}

Multiwavelength photometry from GALEX, SDSS, Gaia, 2MASS, and AllWISE was assembled for 74 catalog sources---43 in the WD locus, 26 in the hot-subdwarf region, and 5 intermediate---of which 57 have no DESI spectrum, so this subsample probes the catalog well beyond the spectroscopic subset. The per-band photometry and the color diagnostics below are released as machine-readable products. \added{Table~\ref{tab:mrt3} shows the first entries of that table, which carries one row for each of the 74 sources and is the data behind Figure~\ref{fig:uvir}.} Figure~\ref{fig:uvir} summarizes two simple color diagnostics.

\textit{Ultraviolet colors.} Thirty-one sources have GALEX NUV detections at S/N$\geq$3, and 14 of these are also detected in the FUV. The WD-locus members are systematically ultraviolet-bright (median ${\rm NUV}-G=+0.8$~mag, versus $+2.3$~mag for the hot-subdwarf-region members, whose kiloparsec distances imply larger foreground extinction), and all FUV detections occur at blue optical colors (Figure~\ref{fig:uvir}(a)). Six sources have ${\rm FUV}-{\rm NUV}<0$, indicating hot photospheric components; the bluest is the optically featureless white dwarf J071816 (Section~\ref{sec:J071816}). Ultraviolet photometry thus identifies hot components independently of the optical spectra, including for sources without any DESI coverage.

\textit{Mid-infrared colors.} Thirty-three sources have AllWISE counterparts within 3~arcsec. $W3$ and $W4$ are never detected (\texttt{ph\_qual}~=~U for every match), so the catalog-level data do not constrain cool dust at 12--22~$\mu$m. $W1$ is detected in all 33 and $W2$ in 18; requiring \texttt{ph\_qual} A or B in both bands, clean contamination flags, and a single blend component leaves 17 sources with reliable $W1-W2$ colors (Figure~\ref{fig:uvir}(b)). Fourteen are consistent with a bare photosphere ($W1-W2\approx0$), as expected in the Rayleigh--Jeans regime. Three show a significant excess ($W1-W2>0.3$~mag at $>2\sigma$): the cataloged cataclysmic variable J214140 ($+0.50\pm0.20$), where disk or companion emission is expected, and two short-period candidates, J194820 ($\Porb=21.95$~min, $+0.92\pm0.23$) and the newly identified J075052 ($\Porb=23.73$~min, $+0.70\pm0.10$). We flag the latter two as candidate composite or accreting systems pending image-level confirmation; the 6-arcsec WISE beam leaves blending possible even for formally clean flags.

\begin{deluxetable*}{lccccccccc}
\tablewidth{0pt}
\tabletypesize{\scriptsize}
\tablecaption{\rev{Machine-readable table of ultraviolet and mid-infrared photometry for the 74-source SED subsample}\label{tab:mrt3}}
\tablehead{
  \colhead{ZTF Name} & \colhead{CMD} & \colhead{FUV} & \colhead{NUV} & \colhead{$G$} & \colhead{$W1$} & \colhead{$W2$} & \colhead{${\rm NUV}-G$} & \colhead{${\rm FUV}-{\rm NUV}$} & \colhead{$W1-W2$} \\
  \colhead{} & \colhead{} & \colhead{(mag)} & \colhead{(mag)} & \colhead{(mag)} & \colhead{(mag)} & \colhead{(mag)} & \colhead{(mag)} & \colhead{(mag)} & \colhead{(mag)}
}
\startdata
ZTFJ000208.56+093543.14 & WD & \nodata & 20.657 & 19.891 & \nodata & \nodata & 0.766 & \nodata & \nodata \\
ZTFJ020052.25-092431.69 & WD & 15.528 & 15.782 & 16.799 & 16.923 & 16.734 & -1.017 & -0.254 & \nodata \\
ZTFJ035352.96+431525.16 & WD & 18.170 & 18.417 & 19.161 & \nodata & \nodata & -0.744 & -0.247 & \nodata \\
ZTFJ045116.84+010426.23 & int & \nodata & \nodata & 15.344 & 14.492 & 14.520 & \nodata & \nodata & -0.028 \\
ZTFJ053433.01+770755.05 & WD & 18.201 & 17.084 & 16.513 & 16.336 & 16.510 & 0.571 & 1.117 & -0.174 \\

\enddata
\tablecomments{\rev{Table~\ref{tab:mrt3} is published in its entirety in machine-readable form. The first five entries with at least one measured color are shown here for guidance regarding its form and content; the machine-readable version carries all 74 SED-subsample sources, the photometric uncertainties, the adopted $\Porb$, the Gaia color, and the $W1-W2$ excess flag. These are the quantities plotted in Figure~\ref{fig:uvir}. Colors are blank where the band is undetected at S/N${}<3$ or where the AllWISE reliability cut fails; $W3$ and $W4$ are never detected and are not tabulated. GALEX magnitudes are AB, AllWISE magnitudes are Vega. The full multiband SED compilation from which these colors are formed is released alongside the table.}}
\end{deluxetable*}

\section{Rotation and Ellipsoidal Diagnostics for the WD-Locus Subsample}
\label{app:rotation}

\added{Table~\ref{tab:rotation} lists, for the 28 WD-locus members with published pure-hydrogen atmospheric fits, the measured peak-to-peak amplitudes and the per-source ellipsoidal and rotation diagnostics discussed in Section~\ref{sec:harmonic_test}. The same quantities are released in machine-readable form with the data products.}

\begin{deluxetable*}{lccccccccc}
\tablewidth{0pt}
\tabletypesize{\scriptsize}
\tablecaption{\rev{Ellipsoidal and rotation diagnostics for the 28 WD-locus members with published pure-hydrogen fits}\label{tab:rotation}}
\tablehead{
  \colhead{Name} & \colhead{$\Pref$} & \colhead{$M_1$} & \colhead{$R_1$} & \colhead{$A_{\rm obs}$} & \colhead{$A_{\rm ellip}^{\rm max}$} & \colhead{$A_{\rm ellip}^{\rm max}/A_{\rm obs}$} & \colhead{$v_{\rm eq}$} & \colhead{$\Pref/P_{\rm crit}$} & \colhead{$fc$} \\
  \colhead{} & \colhead{(min)} & \colhead{($\Msun$)} & \colhead{($R_\odot$)} & \colhead{(mag)} & \colhead{(mag)} & \colhead{} & \colhead{(km\,s$^{-1}$)} & \colhead{} & \colhead{}
}
\startdata
J053332 & 10.2831 & 0.146 & 0.0490 & 0.166 & $4.7\times10^{-2}$ & $2.8\times10^{-1}$ & 347 & 2.2 & 0.142 \\
J213957 & 10.6285 & 0.844 & 0.0097 & 0.101 & $5.8\times10^{-5}$ & $5.8\times10^{-4}$ & 67 & 61 & 0.089 \\
J071816 & 11.2726 & 1.330 & 0.0045 & 0.068 & $3.2\times10^{-6}$ & $4.7\times10^{-5}$ & 29 & 262 & 0.061 \\
J071330$^{\dagger}$ & 11.2801 & 0.651 & 0.0122 & 0.066 & $1.3\times10^{-4}$ & $2.0\times10^{-3}$ & 79 & 40 & 0.059 \\
J164929 & 12.0122 & 0.629 & 0.0125 & 0.059 & $1.3\times10^{-4}$ & $2.2\times10^{-3}$ & 76 & 41 & 0.053 \\
J131845 & 12.1753 & 0.363 & 0.0213 & 0.349 & $1.1\times10^{-3}$ & $3.1\times10^{-3}$ & 128 & 14 & 0.275 \\
J045707 & 16.2500 & 0.585 & 0.0135 & 0.121 & $9.6\times10^{-5}$ & $7.9\times10^{-4}$ & 60 & 48 & 0.106 \\
J211119 & 17.0158 & 0.147 & 0.0673 & 0.063 & $4.3\times10^{-2}$ & $6.9\times10^{-1}$ & 288 & 2.2 & 0.056 \\
J000637 & 23.1538 & 1.095 & 0.0071 & 0.080 & $3.8\times10^{-6}$ & $4.7\times10^{-5}$ & 22 & 242 & 0.071 \\
J162009 & 24.4059 & 0.887 & 0.0092 & 0.133 & $9.1\times10^{-6}$ & $6.8\times10^{-5}$ & 28 & 155 & 0.116 \\
J035352 & 26.4809 & 1.053 & 0.0078 & 0.305 & $3.9\times10^{-6}$ & $1.3\times10^{-5}$ & 21 & 237 & 0.245 \\
J000208 & 28.9522 & 0.824 & 0.0100 & 0.247 & $8.8\times10^{-6}$ & $3.5\times10^{-5}$ & 25 & 158 & 0.203 \\
J035315 & 29.1332 & 1.142 & 0.0066 & 0.215 & $1.8\times10^{-6}$ & $8.3\times10^{-6}$ & 16 & 351 & 0.180 \\
J184326 & 29.6777 & 0.838 & 0.0099 & 0.347 & $8.1\times10^{-6}$ & $2.3\times10^{-5}$ & 24 & 165 & 0.274 \\
J232029 & 34.0557 & 1.017 & 0.0081 & 0.169 & $2.7\times10^{-6}$ & $1.6\times10^{-5}$ & 17 & 282 & 0.144 \\
J003513 & 36.3148 & 0.811 & 0.0101 & 0.086 & $5.9\times10^{-6}$ & $6.8\times10^{-5}$ & 20 & 193 & 0.076 \\
J100316 & 37.5732 & 0.830 & 0.0099 & 0.106 & $5.0\times10^{-6}$ & $4.7\times10^{-5}$ & 19 & 209 & 0.093 \\
J152934 & 38.1466 & 0.959 & 0.0085 & 0.079 & $2.7\times10^{-6}$ & $3.4\times10^{-5}$ & 16 & 285 & 0.070 \\
J200736 & 40.8153 & 0.464 & 0.0153 & 0.141 & $2.8\times10^{-5}$ & $2.0\times10^{-4}$ & 27 & 88 & 0.122 \\
J055805 & 43.0406 & 0.261 & 0.0264 & 0.122 & $2.3\times10^{-4}$ & $1.9\times10^{-3}$ & 45 & 31 & 0.106 \\
J053433 & 43.4079 & 0.845 & 0.0098 & 0.167 & $3.6\times10^{-6}$ & $2.1\times10^{-5}$ & 16 & 247 & 0.142 \\
J110045 & 48.1550 & 0.275 & 0.0276 & 0.120 & $2.0\times10^{-4}$ & $1.7\times10^{-3}$ & 42 & 33 & 0.104 \\
J161133 & 49.7082 & 0.296 & 0.0220 & 0.087 & $3.5\times10^{-4}$ & $4.1\times10^{-3}$ & 32 & 50 & 0.077 \\
J060914$^{\dagger}$ & 51.1756 & 0.242 & 0.0324 & 0.098 & $3.3\times10^{-4}$ & $3.3\times10^{-3}$ & 46 & 26 & 0.086 \\
J160335 & 78.3939 & 0.426 & 0.0167 & 0.203 & $4.3\times10^{-5}$ & $2.1\times10^{-4}$ & 16 & 142 & 0.170 \\
J214140 & 78.7562 & 0.605 & 0.0129 & 0.282 & $1.4\times10^{-5}$ & $4.9\times10^{-5}$ & 12 & 251 & 0.229 \\
J103533 & 82.0896 & 0.786 & 0.0105 & 0.256 & $5.3\times10^{-6}$ & $2.1\times10^{-5}$ & 9 & 407 & 0.210 \\
J190409 & 88.1386 & 0.602 & 0.0128 & 0.226 & $1.1\times10^{-5}$ & $4.8\times10^{-5}$ & 11 & 283 & 0.188 \\

\enddata
\tablecomments{\rev{Ordered by $\Pref$. $M_1$ and $R_1$ are the pure-hydrogen photometric values of \citet{GentileFusillo2021}. $A_{\rm obs}$ is the larger of the $g$- and $r$-band peak-to-peak phase-binned amplitudes (Section~\ref{sec:individual}). A dagger marks the two sources whose folded profiles do not reproduce between the two halves of the ZTF baseline; they are listed for completeness but excluded from the statistics quoted in the text. $A_{\rm ellip}^{\rm max}$ is the maximum peak-to-peak ellipsoidal amplitude at the adopted $\Porb$ for an equal-mass companion and an edge-on orbit. The last three columns evaluate the rotation hypothesis at $P_{\rm rot}=\Pref$: the equatorial velocity, the ratio to the Keplerian break-up period $P_{\rm crit}=2\pi(R_1^{3}/GM_1)^{1/2}$, and the projected covering fraction times band contrast, $fc=1-10^{-0.4A_{\rm obs}}$, that a rotating surface feature would need. J053332 (PTF~J0533+0209) and J211119 are the two extremely low-mass members for which the ellipsoidal interpretation remains viable and the rotational one does not.}}
\end{deluxetable*}

\section{Gravitational-Wave Signal Model}
\label{app:gwmodel}

\subsubsection{Gravitational Wave Signal Model}

We follow the formalism of \citet{Yu2026} and treat each signal as monochromatic over the mission lifetime. In this approximation, the two GW polarization modes in the principal polarization frame are:
\beq
h_+(t) = A\,(1+\cos^2\iota)\,\cos\!\bigl[2\pi\fGW\,t + \phi_0 + \Phi_D(t)\bigr],
\label{eq:hplus}
\eeq
\beq
h_\times(t) = 2A\cos\iota\,\sin\!\bigl[2\pi\fGW\,t + \phi_0 + \Phi_D(t)\bigr],
\label{eq:hcross}
\eeq
where $\iota$ is the orbital inclination, $\phi_0$ is the initial phase, and the Doppler modulation phase arising from the detector's annual motion around the Sun is
\beq
\Phi_D(t) = \frac{2\pi\fGW R_\oplus}{c}\,\sin\!\left(\frac{\pi}{2}-\beta\right)\cos(2\pi f_m t - \lambda),
\eeq
with $R_\oplus = 1$~AU, $f_m = 1$~yr$^{-1}$, and $(\lambda, \beta)$ the ecliptic longitude and latitude of the source.  The dimensionless GW strain amplitude is \citep{Peters1963, Flanagan1998}:
\beq
A = \frac{2(G\Mc)^{5/3}}{c^4\,d}\,(\pi\fGW)^{2/3},
\label{eq:strain_A}
\eeq
where $\Mc = (m_1 m_2)^{3/5}/(m_1+m_2)^{1/5}$ is the chirp mass and $d$ is the luminosity distance.  The sky-, polarization- and inclination-averaged strain commonly quoted in the literature \citep{Robson2019} is
\beq
h = \frac{8}{\sqrt{5}}\,\frac{(G\Mc)^{5/3}}{c^4\,d}\,(\pi\fGW)^{2/3}
  = \frac{4}{\sqrt{5}}\,A,
\label{eq:strain}
\eeq
where the factor $4/\sqrt{5}$ comes from inclination, sky, and polarization averaging \citep{Robson2019}. Throughout this section we use Equation~(\ref{eq:strain}) with sky-averaged sensitivity curves $S_n(f)$, so that the characteristic strain $h_c=h\sqrt{N_{\rm cyc}}$ enters the matched-filter SNR. The recovered verification binary ES~Cet anchors the normalization against the known verification-binary population (Figure~\ref{fig:gw_strain}).

\subsubsection{Signal-to-Noise Ratio}

The matched-filter SNR for a monochromatic signal \citep{Yu2026} is
\beq
\rho^2 = (h|h) \approx \frac{2}{S_n(\fGW)}\int_0^T h^2(t)\,dt,
\label{eq:snr}
\eeq
where $S_n(f)$ is the one-sided power spectral density (PSD) of the detector noise and $T$ is the observation time.  Equivalently, the SNR can be estimated from the ratio of the characteristic strain $h_c = h\sqrt{N_{\rm cyc}}$ to the noise characteristic strain $h_n(f) = \sqrt{f S_n(f)}$, where $N_{\rm cyc} = \fGW\,T$ is the number of GW cycles accumulated during the mission.  For a network of $n$ independent detectors, the combined SNR is $\rho_{\rm net}^2 = \sum_i \rho_i^2$ \citep{Yu2026}.

We adopt the sensitivity curves of \citet{Yu2026} for four space-based detectors: TianQin (5~yr with a 50\% duty-cycle factor) and 4-yr effective integrations for LISA, Taiji, and DECIGO. Key parameters are arm lengths $L_{\rm TQ} = \sqrt{3}\times10^8$~m, $L_{\rm LISA} = 2.5\times10^9$~m, $L_{\rm Taiji} = 3\times10^9$~m, and $L_{\rm DECIGO} = 10^6$~m; displacement noise floors $S_x$ and acceleration noise levels $S_a$ are taken from Table~I of \citet{Yu2026}.  For LISA and Taiji, which observe the mHz band where the unresolved Galactic double-white-dwarf foreground dominates, we add the 4-yr Galactic confusion noise of \citet{Robson2019} (their Eq.~14) to the instrument PSD; TianQin (short arm, higher-frequency band) and DECIGO (deci-Hz) are treated as instrument-noise-limited in this band. The confusion term is dominant over the LISA/Taiji instrument noise below $\sim$1~mHz and therefore materially affects the ranking of our lowest-frequency candidates.

\begin{deluxetable*}{lccccccc}
\tablewidth{0pt}
\tabletypesize{\footnotesize}
\tablecaption{Fiducial GW estimates for WD cooling-sequence (WD-CS) candidates and a literature benchmark\label{tab:gw}}
\tablehead{
  \colhead{ZTF Name} & \colhead{Locus} & \colhead{Type} & \colhead{Assumed $\Porb$} & \colhead{Assumed $\fGW$} & \colhead{$d$} &
  \colhead{$\rho_{\rm TQ}$} & \colhead{$\rho_{\rm LISA}$} \\
  \colhead{} & \colhead{} & \colhead{} & \colhead{(min)} & \colhead{(mHz)} & \colhead{(pc)} &
  \colhead{} & \colhead{}
}
\startdata
J071816.38$+$373138.66 & WD-CS & EW & 22.5452 & 1.479 & 85 & 79.3 & 220.5 \\
J020052.25$-$092431.69\tablenotemark{a} & special & EW & 10.3369 & 3.225 & 1787 & 29.6 & 120.8 \\
J162009.42$+$125647.33 & WD-CS & EW & 48.8117 & 0.683 & 75 & 11.5 & 29.3 \\
J000637.94$+$310415.53 & WD-CS & EW & 46.3076 & 0.720 & 98 & 10.1 & 25.4 \\
J152934.91$+$292801.87 & WD-CS & EW & 76.2931 & 0.437 & 87 & 3.0 & 8.8 \\
J100316.62$+$354354.08 & WD-CS & EW & 75.1464 & 0.444 & 107 & 2.5 & 7.4 \\
J000208.56$+$093543.14 & WD-CS & EW & 57.9044 & 0.576 & 249 & 2.2 & 5.9 \\
J103533.01$+$055159.00 & WD-CS & EA & 82.0896 & 0.406 & 197 & 1.1 & 3.2 \\
J160335.93$+$215032.33 & WD-CS & EA & 78.3939 & 0.425 & 308 & 0.8 & 2.3 \\

\enddata
\tablecomments{Values assume $\Mc=0.3\,\Msun$ and $\fGW=2/\Porb$. Except for ES~Cet, $\Porb$ is an adopted orbital-period estimate inferred from the source-by-source photometric harmonic; the tabulated SNRs are fiducial values under the assumptions of Section~\ref{sec:gw}. \added{Eight of the nine DESI-characterized WD cooling-sequence members are listed: the ninth, J214140, is one of the three multiband-poor sources whose adopted periods Section~\ref{sec:period_def} flags as unreliable, and it is excluded from the GW ranking for that reason. These rows are an excerpt of the machine-readable table of fiducial gravitational-wave quantities, which is published in its entirety for all 147 candidates.}}
\tablenotetext{a}{ES~Cet; uses the independently established orbit, $\Porb=10.3369$~min.}
\end{deluxetable*}

\end{document}